\documentclass[A4,12pt,fleqn]{article}

\usepackage{amsmath,cite,epsfig,multicol}
\usepackage{color}
\usepackage{bbm}
\usepackage{lscape}

\def\b               {\beta}

\def\t               {\theta}

\def\x               {\chi}

\def\L               {{\cal L}}

\def\P               {{\cal P}}

\def\ti              {\tilde}

\def\nt              {\ti\x^0}
\def\ch              {\ti\x^\pm}
\def\chm              {\ti\x^-}
\def\sf              {\ti f}

\def\stop            {\ti t}
\def\sbottom         {\ti b}

\def\tsf             {\theta_{\ti f}}
\def\sb         {\ti b}

\def\bsg{{\rm BR}(B\to X_s\gamma)}
\def\bmm{{\rm BR}(B_s\to\mu^+\mu^-)}
\def\bsgsub{{\rm BR}_{B\to X_s\gamma}}
\def\bmmsub{{\rm BR}_{B_s\to\mu^+\mu^-}}

\def\Pt {\P_{t}^{}}

\newcommand{\mnt}[1] {m_{\ti \x^0_{#1}}}
\newcommand{\mch}[1] {m_{\ti \x^\pm_{#1}}}
\newcommand{\msf}[1] {m_{\sf_{#1}}}

\newcommand{\msb}[1] {m_{\ti b_{#1}}}

\def\PL              {P_L^{}}
\def\PR              {P_R^{}}

\def\to              {\rightarrow}

\newcommand{\nn}{\nonumber}
\newcommand{\noi}{\noindent}

\newcommand{\mbf}      {\boldmath}
\newcommand{\sfrac}[2] {{\textstyle \frac{#1}{#2}}}

\newcommand{\eq}[1]  {\mbox{(\ref{eq:#1})}}

\newcommand{\ie}{{\it i.e.}}

\newcommand{\lsim}{\;\raisebox{-0.9ex}{$\textstyle\stackrel{\textstyle<}
           {\sim}$}\;}

\newcommand{\comment}[1]{}

\begin{document}
\vspace*{-18mm}
\begin{flushright}
LAPTH-019-13\\
LSPSC13094
\end{flushright}

\vspace*{6mm}

\begin{center}

{\LARGE\bf Top Polarization in Sbottom Decays \\[3mm] at the LHC}

\vspace*{1cm}

{\large Genevi\`eve~B\'elanger${^1}$, Rohini~M.~Godbole${^2}$,
        Sabine~Kraml${^3}$, Suchita~Kulkarni${^3}$}

\vspace*{1cm}

{\it
$^1$~LAPTH, Universit\'e de Savoie, CNRS, B.P.110, F-74941 Annecy-le-Vieux Cedex, France\\[2mm]
$^2$~Centre for High Energy Physics, Indian Institute of Science,
     Bangalore 560012, India\\[2mm]
$^3$~Laboratoire de Physique Subatomique et de Cosmologie, UJF Grenoble 1, 
CNRS/IN2P3, INPG, 53 Avenue des Martyrs, F-38026 Grenoble, France\\[2mm]
}

\end{center}

\vspace*{1cm}

\begin{abstract}
\noindent
We perform a comprehensive analysis of the polarization of the top quarks originating from sbottom-pair production 
at the LHC, followed by sbottom decays to top+chargino. We study moreover the expected net  polarization 
of top quarks produced in  sbottom-to-chargino and stop-to-neutralino decays in scenarios with small 
chargino--neutralino mass difference, where these decays may be hard to distinguish.
We show that, in contrast to top quarks produced via Standard Model processes,  
the average polarization of top quarks originating from these SUSY decays can obtain any 
value between $+1$ and $-1$. 
We further study the effect of this polarization on the top quark decay kinematics. 
On the one hand this may be used to 
construct measures of this polarization, on the other hand it may be used to enhance the search reach 
in certain scenarios. 
Exploiting top polarization may also prove useful for searches for ``natural'' SUSY with light higgsinos, 
which is typically very difficult to detect at the LHC.
\end{abstract}

\clearpage
\section{Introduction}

The discovery of the Higgs boson  by both the ATLAS and CMS 
collaborations~\cite{Aad:2012tfa,Chatrchyan:2012ufa} has provided the last missing piece of the 
Standard Model (SM). Nevertheless there are still fundamental problems open in the SM,  such as the 
nature of dark matter, the origin of CP violation, or the stability of the electroweak scale, 
motivating the need for new physics beyond the SM. Supersymmetry (SUSY) is one of the best-motivated 
extensions of the SM for addressing some of these issues, and the search for supersymmetric particles 
is thus one of the primary objectives of LHC experiments. 

While SUSY searches at the LHC with $\sqrt{s}=7$--8~TeV have pushed the mass limits for 
gluinos and light-flavor squarks beyond 1~TeV, 
current limits on third generation squarks are much weaker for several reasons.
First, the cross section for direct pair production of stops and sbottoms is dominated by gluon-initiated 
processes and thus drops rapidly as their mass increases. Second, this cross section is much smaller 
than the total squark--gluino cross section that contributes to the production of SUSY partners of light quarks. 
Finally, the decays of stops and sbottoms involve some final states with top quarks, which means that the 
results from the generic $E_T^{\rm miss}$+jets searches are not applicable. 

Scenarios with an inverted mass hierarchy, {\it i.e.}\ light third but heavy 1st/2nd generation squarks, 
have hence become a new focus of phenomenological studies. In particular it is interesting to investigate 
specific  signatures of  stops and sbottoms  as well as methods  to determine their properties at the LHC.

Results on stop and sbottom searches in direct pair production are usually quoted in terms of 
Simplified Model Spectra (SMS), which assume  100\% branching ratio in a given channel.  
The best limit on stops are currently  obtained in the $\tilde t \rightarrow t \tilde\chi^0_1$ channel: 
based on ${\cal L}=21~{\rm fb}^{-1}$ of data at 8~TeV, the ATLAS collaboration has excluded stops 
in the mass range 320--680~GeV in the limit of massless neutralinos~\cite{ATLAS-stop}. 
Slightly weaker upper limits (600 GeV) are obtained from the channel $\tilde t \rightarrow b \tilde\chi^+_1$ 
assuming a small mass difference between the chargino and the neutralino LSP~\cite{ATLAS-stop-bchargino}, 
while sbottom masses below 620~GeV are excluded in the channel $\sbottom\to b\tilde\chi^0_1$  for neutralino 
masses below 150~GeV~\cite{ATLAS-sbottom} (both results were obtained by the ATLAS collaboration with ${\cal L}=12.8~{\rm fb}^{-1}$). 
These limits become much weaker with smaller mass splittings between the squark and the neutralino/chargino. 
Present data therefore allow both top and bottom squarks well below the TeV scale. 

For final states involving top quarks, the top polarization can be a useful tool to probe new physics 
at colliders, as it is sensitive to the helicity structure of the production process 
(for a recent summary see {\it e.g.}~Ref.~\cite{Godbole:2013ya}).
There are different ways to measure the top polarization.
In particular, there is a strong correlation between the polarization of the top quark and the angular 
distributions of its decay leptons. It was shown that this correlation is not affected by higher-order 
corrections~\cite{Jezabek:1988ja,Czarnecki:1990pe,Brandenburg:2002xr} or new physics 
contributions to the 
decay~\cite{Grzadkowski:2001tq,Grzadkowski:2002gt,Godbole:2002qu,Godbole:2006tq,Godbole:2010kr}.
Measures of top polarization  using angular variables have been 
constructed~\cite{Godbole:2006tq,Godbole:2010kr}
The energy fraction of the decay leptons may also be exploited as a measure of the top polarization~\cite{Berger:2012an}.  
Moreover, it was suggested to make use of boosted top and jet substructure methods in hadronic decays to determine the top polarization~\cite{Krohn:2009wm}. Finally, the determination of the top polarization at the LHC in $t\bar{t}$ events using the angular distributions of the top decay products in the top rest frame was investigated in~\cite{ATLAS-top-pol}.

In the SM, pair-produced tops are unpolarized, while singly produced tops have polarization $-1$.  
In SUSY, the polarization of top quarks produced in the decays of stops or sbottoms can take any 
value between $\pm1$, and may provide information on the underlying SUSY scenario. 
In the context of $e^+e^-$ colliders, 
the top polarization may be used to probe the mixing in the stop sector~\cite{Boos:2003vf,Gajdosik:2004ed}. In this case, one can also use the production cross section which depends on the squark mixing. At the LHC, the production cross section is independent of the mixing angle, leaving mainly the polarization to extract information on the stop/sbottom mixing (in addition to branching ratios, if they can be extracted). 

Possibilities of measuring the top polarization and hence getting information on the stop mixing in direct stop-pair 
production at the LHC with $\sqrt{s}=14$~TeV were investigated in~\cite{Perelstein:2008zt,Bhattacherjee:2012ir}. 
More generally, in~\cite{Belanger:2012tm} some of us showed how the longitudinal polarization of the top quark 
from stop decays into neutralinos, $\tilde t_1\rightarrow t \tilde\chi^0_{1,2}$, depend on the mixing in both the stop and the neutralino sectors, as well as on the mass difference between the stop and the neutralino. 
More precisely, for large mass difference $\Delta m=m_{\stop}-m_{\tilde\chi^0}-m_t$, a right-handed (RH) stop produces a negative top polarization when it decays into a higgsino and a positive polarization when it decays into a bino, and vice-versa for a left-handed (LH) stop. 
On the other hand, the top polarization vanishes in the limit of small $\Delta m$.
Furthermore, as pointed out in \cite{Belanger:2012tm,Low:2013aza}, 
it will  affect observables such as the energy distribution of the lepton resulting from the top decay, thus  impacting the reach for stop searches at the LHC. 
In fact, different assumptions about the top polarization are at least in part responsible for the higher reach 
in the stop mass exclusion by ATLAS (which assumes RH top quarks) than by CMS 
(which assumes unpolarized top quarks)~\cite{CMS-SUS-023,CMS-stop-talk}.
Similar effects are expected for sbottom searches in the top+chargino decay channel.

In this paper, we extend the study of \cite{Belanger:2012tm} by 
investigating the behavior of the top polarization in sbottom decays into charginos,
$${\tilde b}\rightarrow t \tilde\chi^- \,.$$ 
As in the case of stop decays, this polarization may give information on the nature of both the sbottom 
and the chargino. 
We will show that for a wino-like chargino the polarization of the top is always $-1$, while for a higgsino-like chargino the polarization varies from $-1$ for a RH sbottom to $+1$ to a LH sbottom.  Thus a positive polarization would give indication on the higgsino nature of the chargino. Recall that higgsino-like charginos and neutralinos have small direct production at the LHC and are therefore hard to probe directly especially since the decay products are soft because of the small mass difference between the higgsino states. 
In reality, the situation is more complicated: 
when the lighter sbottom $\sbottom_1$ has a large LH component, its mass is similar to that of the lighter stop $\stop_1$ 
(as both are determined by the same soft mass parameter), and therefore tops can come from several different channels, 
${\tilde b}\rightarrow t \tilde\chi^+$, $\tilde t \rightarrow t\tilde\chi^0$. At the end, only the net polarization resulting from all 
different decays will be measurable. Relating the top polarization to the underlying properties of stops, sbottoms and neutralinos/charginos  therefore becomes more challenging.

The  paper is organized as follows. We summarize our notation and conventions in Section~2 before discussing the fermion polarization in the squark rest frame in Section 3. Polarization-dependent observables are presented in Section~4, and the treatment of the boost in Section~5. Section~6 contains all numerical results, including the dependence of the polarization on the fundamental SUSY parameters,  some benchmarks scenarios corresponding to different net polarizations, as well as an analysis of polarization-dependent observables that are relevant for searches at the 14~TeV LHC. Section~7 contains our conclusions.

\section{Notation and conventions}

For completeness we review in this section  our notation and conventions, 
and give the relevant expressions for sfermion decays into charginos and neutralinos. 
Overall, we follow the notation of \cite{Gajdosik:2004ed}, which was also used in \cite{Godbole:2011vw}. 

\subsection{Sfermion system}

Ignoring intergenerational mixing, the sfermion mass matrices can
be written as a series of $2\!\times\!2$ matrices, each of which
describes sfermions of a specific flavour:
\begin{equation}
  {\cal M}_{\sf}^2 =
  \left( \begin{array}{cc}  \msf{L}^2 & a_f m_f \\
                            a_f m_f   & \msf{R}^2
  \end{array}\right) \;=\;
  (R^{\sf})^T \left( \begin{array}{cc} \msf{1}^2 & 0 \\
                                            0         & \msf{2}^2
                   \end{array}\right) R^{\sf}
\label{eq:msfmat}
\end{equation}
with
\begin{eqnarray}
  \msf{L}^2 &=& M^2_{\ti L}
    + m_Z^2 \cos 2\beta\,(I_{3L}^f - e_f\sin^2\t_W) + m_f^2, \nonumber\\[2mm]
  \msf{R}^2 &=& M^2_{\ti R}
    + e_f\,m_Z^2 \cos 2\b\,\sin^2\t_W + m_f^2, \nonumber\\[1mm]
  a_f &=& A_f - \mu\, \{ \cot\b , \tan\b \}\,,
  \label{eq:aq}
\end{eqnarray}
for $\{$up, down$\}$-type sfermions; $m_f$, $e_f$ and $I_{3}^f$
are the mass, electric charge and the third component of the weak
isospin of the partner fermion, respectively; $M_{\ti L}$,
$M_{\ti R}$ and $A_f$ are soft SUSY-breaking parameters for each
family, and $\mu$ is the higgsino mass parameter.\\

\noi
According to eq.~\eq{msfmat}, ${\cal M}_{\sf}^2$ is diagonalized by a
unitary rotation matrix $R^{\sf}$.
The weak eigenstates $\sf_L^{}$ and $\sf_R^{}$ are thus
related to their mass eigenstates $\sf_1^{}$ and $\sf_2^{}$ by
\begin{equation}
  {\sf_1^{} \choose \sf_2^{}} = R^{\sf}\,{\sf_L^{} \choose \sf_R^{}},
  \hspace{8mm}
  R^{\sf} = \left(\begin{array}{cc}
                     \cos\tsf\; & \sin\tsf \\
                    -\sin\tsf             & \cos\tsf\;
            \end{array}\right) .
\label{eq:Rsf}
\end{equation}
Since the off-diagonal element of ${\cal M}_{\sf}^2$ is
proportional to $m_f$, this mixing is mostly relevant to the third
generation, $\ti f=\ti t,\ti b$, on which we concentrate
in the following. The mass eigenvalues are given by
\begin{equation}
  m^2_{\sf_{1\!,2}} = \frac{1}{2} \left( \msf{L}^2 + \msf{R}^2
  \mp \sqrt{(\msf{L}^2 - \msf{R}^2)^2 + 4\, (a_f m_f)^2 } \,\right).
\label{eq:sfmasseigenvalues}
\end{equation}
By convention, we choose $\sf_1^{}$ to be the lighter mass
eigenstate, $\msf{1}\leq\msf{2}$. Notice also that
$\msf{1}\leq\msf{L,R}\leq\msf{2}$. For the mixing angle $\tsf$ we
choose
\begin{equation}
  \cos\tsf = \frac{-a_f\,m_f}
                  {\sqrt{(\msf{L}^2-\msf{1}^2)^2 + (a_f m_f)^2}} \,,
  \qquad
  \sin\tsf = \frac{\msf{L}^2-\msf{1}^2}
                  {\sqrt{(\msf{L}^2-\msf{1}^2)^2 + (a_f m_f)^2}} \,.
\end{equation}
The $\sf_L$--$\sf_R$ mixing is large if $(\msf{L}^2 - \msf{R}^2) \lsim
(a_f m_f)$, with $|\cos\tsf | > \frac{1}{\sqrt 2}\,$ if
$\msf{L}<\msf{R}$ and $| \cos\tsf | < \frac{1}{\sqrt 2}\,$ if
$\msf{R}<\msf{L}$.

\subsection{Neutralino system}

In the basis
\begin{equation}
  \Psi_j^0=\left(-i\lambda ',-i\lambda^3,\psi_{H_1}^0,\psi_{H_2}^0\right)
\end{equation}
the neutralino mass matrix is:
\begin{equation}
  {\cal M}_N =
  \left( \begin{array}{cccc}
  M_1 & 0 & -m_Z\sin\theta_W\cos\beta  & m_Z\sin\theta_W\sin\beta \\
  0 & M_2 &  m_Z\cos\theta_W\cos\beta  & -m_Z\cos\theta_W\sin\beta  \\
  -m_Z\sin\theta_W\cos\beta & m_Z\cos\theta_W\cos\beta   & 0 & -\mu \\
   m_Z\sin\theta_W\sin\beta & - m_Z\cos\theta_W\sin\beta & -\mu & 0
  \end{array}\right) .
\label{eq:ntmassmat}
\end{equation}

\noi The matrix of eq.~\eq{ntmassmat} is diagonalized by the
unitary mixing matrix $N$:
\begin{equation}
  N{{\cal M}_N} N^T =
  {\rm diag}(\mnt{1},\,\mnt{2},\,\mnt{3},\,\mnt{4})\,,
\end{equation}
where $\mnt{n}$, $n=1,...,4$, are the (non-negative) masses of the
physical neutralino states with $\mnt{1}<....<\mnt{4}$. 

\subsection{Chargino system}

\noi
The chargino mass matrix is:
\begin{equation}
  {\cal M}_C =
  \left( \begin{array}{cc}
    M_2 &\sqrt 2\, m_W\sin\beta \\
    \sqrt 2\,m_W\cos\beta & \mu
  \end{array}\right) \,.
\end{equation}
It is diagonalized by  two unitary matrices $U$ and $V$,
\begin{equation}
  U{\cal M}_C V^T = {\rm diag}(\mch{1},\,\mch{2})\,,
\end{equation}
where $\mch{1,2}$ are the masses of the physical chargino states
with $\mch{1}<\mch{2}$.

\subsection{Sfermion interaction with charginos and neutralinos} 

The sfermion interactions with charginos, which will define the $\sbottom \to t\chm $ decays,
are %
\begin{eqnarray}
  \L_{f'\!\sf\ch}
  &=& g\,\bar u\,( l_{ij}^{\,\ti d}\,\PR +
                   k_{ij}^{\,\ti d}\,\PL )\,\ti\x^+_j\,\ti d_i^{}
    + g\,\bar d\,( l_{ij}^{\,\ti u}\,\PR +
                   k_{ij}^{\,\ti u}\,\PL )\,\ti\x^{+c}_j\,\ti u_i^{}
      + {\rm h.c.}
\end{eqnarray}
where  $i,j=1,2$, $u$ ($\ti u$) stands for up-type (s)quark,
and $d$ ($\ti d$) stands for down-type (s)quark, $g$ is the SU(2) coupling constant.
The coupling matrices  $l^{\ti f}$ and $k^{\ti f}$ are
\begin{align}
  l_{ij}^{\ti t} &= -V_{j1} R_{i1}^{\ti t} + h_t\,V_{j2} R_{i2}^{\ti t}\,, &
  l_{ij}^{\ti b} &= -U_{j1} R_{i1}^{\ti b} + h_b\,U_{j2} R_{i2}^{\ti b}\,,
  \label{eq:elltb} \\
  k_{ij}^{\ti t} &= h_b\,U_{j2} R_{i1}^{\ti t} \,, &
  k_{ij}^{\ti b} &= h_t\,V_{j2} R_{i1}^{\ti b} \,,
  \label{eq:katb}
\end{align}
for stops and sbottoms, with the Yukawa couplings $h_f$ given by
\begin{equation}
  h_t=\frac{m_t}{\sqrt 2\,m_W\sin\b}\,,\qquad
  h_b=\frac{m_b}{\sqrt 2\,m_W\cos\b}\,.
  \label{eq:yuk}
\end{equation}

The sfermion interactions with neutralinos are ($i=1,2$; $n=1,...,4$)
\begin{eqnarray}
  \L_{f\sf\nt}
  &=& g\,\bar f\,( f_{Ln}^{\sf}\PR + h_{Ln}^{\sf}\PL )\,\nt_n\,\sf_L^{} +
      g\,\bar f\,( h_{Rn}^{\sf}\PR + f_{Rn}^{\sf}\PL )\,\nt_n\,\sf_R^{}
      + {\rm h.c.}\nn\\
  &=& g\,\bar f\,( a^{\,\sf}_{in}\PR + b^{\,\sf}_{in}\PL )\,\nt_n\,\sf_i^{}
      + {\rm h.c.}
\end{eqnarray}
where
\begin{eqnarray}
   a^{\,\sf}_{in} &=& f_{Ln}^{\sf}\,R_{i1}^{\sf} +
                      h_{Rn}^{\sf}\,R_{i2}^{\sf},   \label{eq:aik}\\
   b^{\,\sf}_{in} &=& h_{Ln}^{\sf}\,R_{i1}^{\sf} +
                      f_{Rn}^{\sf}\,R_{i2}^{\sf}.   \label{eq:bik}
\end{eqnarray}
The $f_{L,R}^{\sf}$ and $h_{L,R}^{\sf}$ couplings are
\begin{align}
  f_{Ln}^{\,\ti t} &= -\sfrac{1}{\sqrt 2}\,(N_{n2}
                        +\sfrac{1}{3}\tan\theta_W N_{n1}) \,, &
  f_{Ln}^{\,\ti b} &= \sfrac{1}{\sqrt 2}\,(N_{n2}
                        -\sfrac{1}{3}\tan\theta_W N_{n1}) \,, \nonumber\\
  f_{Rn}^{\,\ti t} &= \sfrac{2\sqrt 2}{3}\,\tan\theta_W N_{n1}\,, &
  f_{Rn}^{\,\ti b} &= -\sfrac{\sqrt 2}{3}\,\tan\theta_W N_{n1}\,, \nonumber\\
  h_{Rn}^{\ti t} &= -h_t\, N_{n4} = h_{Ln}^{\ti t*}\,, &
  h_{Rn}^{\ti b} &= -h_b\, N_{n3} = h_{Ln}^{\ti b*}\,
\end{align}
for stops ($\sf=\stop$) on the left and sbottoms ($\sf=\sbottom$) on the right.

\section{\mbf Fermion polarization in the sfermion rest frame} \label{sect:Pfsfrest} 

For a sfermion decaying into a chargino or a neutralino, the gaugino interaction conserves the helicity 
of the sfermion while the higgsino interaction flips it.  
We define the average polarisation of the produced fermions as 
\begin{equation}
P_f=\frac{\sigma(+,+)-\sigma(-,-)}{\sigma(+,+)+\sigma(-,-)},
\label{eq:Ptdef}
\end{equation}
where $\sigma(\pm,\pm)$ is the cross section for a positive or 
negative helicity fermion respectively. In general one expects 
non trivial polarization effects only for the third generation fermions, furthermore  only  the polarization of the top quark or the $\tau$ can be measured. In this paper we will discuss only the top polarization.

\subsection{\mbf $\sbottom \to t \chm$ decays} \label{sect:Pfpr}

For tops coming from $\sbottom_i\to t\,\chm_j$ decays, the polarisation 
given by eq.~\eq{Ptdef}, is 
\begin{equation}
      \P_t^{}  = \frac{ \big[\, {(k_{ij}^{\sbottom})}^2 - {(l_{ij}^{\,\sbottom})}^2\, \big] \,f_1^{} }
                  { {(k_{ij}^{\sbottom})}^2 + {(l_{ij}^{\,\sbottom})}^2 - 2\,k_{ij}^{\sbottom}\,l_{ij}^{\,\sbottom}\,f_2^{} } \,,
\label{eq:Pfprime}
\end{equation}
with the factors $f_1$ and $f_2$,
purely kinematical in origin, given by
\begin{equation}
  f_1\ =\ m_t\, \frac{(p_{\chm_j}\,\cdot\, s_t)}{(p_t\,\cdot\, p_{\chm_j})} \, ,
\qquad
  f_2\ =\ m_t\, \frac{m_{\chm_j}} {(p_t\,\cdot\, p_{\chm_j})} \, .
\label{eq:kinegeneral}
\end{equation}
Here $m_t$, $p_t$ and $s_t$ denote the top mass, momentum and longitudinal spin vector, respectively, 
and $p_{\chm_j}$ is the momentum of the chargino. 
In the rest frame of the decaying sbottom, these factors become
\begin{equation}
  f_1  =  \frac{
  \lambda^{\frac{1}{2}}(m_{\sbottom_i}^2,m_{t}^2,m_{\chm_j}^2)}
     {m_{\sbottom_i}^2-m_{\chm_j}^2-m_{t}^2} \,, \qquad
  f_2 = \frac{2 m_{t} m_{\chm_j}} {m_{\sbottom_i}^2-m_{\chm_j}^2-m_{t}^2} \, ,
  \label{eq:kinerestframe}
\end{equation}
with the function $\lambda$ defined as 
\begin{equation}
\lambda(x,y,z) = x^2+y^2+z^2-2\,x\,y-2\,y\,z-2\,x\,z.
\end{equation}
The factors $f_1\to 1$ and $f_2\to 0$ if $m_t$ was negligible.
Moreover, for large $\Delta m\equiv m_{\sbottom_i}-m_{\chm_j}-m_{t}$,  $f_1\to 1$ irrespective of $m_t$. 
For $\sbottom_1\to t \chm$ decays, we have 
\begin{eqnarray}
  {(k_{1j}^{\ti b})}^{2} - {(l_{1j}^{\ti b})}^{2}
  &=&  h_t^2 V_{j2}^2  \cos^2\theta_{\sbottom} - ( h_b U_{j2}\sin\theta_{\sbottom} - U_{j1}\cos\theta_{\sbottom})^{2}  \nn\\
  &=&  ( h_t^2 V_{j2}^2 - U_{j1}^2 ) \cos^2\theta_{\sbottom}
           - h_b^2 U_{j2}^2 \sin^2\theta_{\sbottom}
           +\, h_b \,U_{j1} U_{j2}  \sin 2\theta_{\sbottom} \,.
\label{eq:sbot1}
\end{eqnarray}
For $\ti b_2^{}$ decays, the corresponding expression  ${(k_{2j}^{\ti b})}^{2} - {(l_{2j}^{\ti b})}^{2}$
is given by the RHS of eq.~\eq{sbot1} with $\cos^2\theta_{\sbottom}$, $\sin^2\theta_{\sbottom}$
interchanged, and a change in sign of the term $\propto\sin 2\theta_{\sbottom}$. 

It is interesting to consider certain limiting cases for the $\ti b_1\to t\ti\x^-_1$ decay. 
For $M_2\ll |\mu|$, the chargino is {\bf wino-like}, and the mixing matrices $U$ and $V$ are given by:\footnote{Up to an overall phase.}
\begin{equation}
U\rightarrow \left( \begin{array}{ccc}
-1 & 0\\
0 & 1\end{array} \right) \,, \qquad
V\rightarrow \left( \begin{array}{ccc}
-1 & 0\\
0 & 1\end{array} \right) \,.
\end{equation}
The resulting top polarization $\Pt$ is then: 
\begin{equation}
   \Pt  = \frac{-{(l_{11}^{\,\sbottom})}^2\,f_1}
                  {{(l_{11}^{\,\sbottom})}^2}  = - f_1 \,.
\end{equation}
Therefore, for large enough $\Delta m$, we expect $\Pt\simeq -1$ in case of a pure wino-like chargino, irrespective of the sbottom mixing angle. (It should be noted however that in practice for the range of parameters we will be considering all  entries of the matrices $U$ and $V$ are non-zero and will be taken into account.)\\ 

\noi For a  {\bf higgsino-like} chargino, i.e.\ $|\mu|\ll M_2$, the mixing matrices $U$ and $V$ approach
\begin{equation}
U\rightarrow \left( \begin{array}{ccc}
0 & 1\\
1 & 0\end{array} \right)\,, \qquad
V\rightarrow \left( \begin{array}{ccc}
0 & 1\\
1 & 0\end{array} \right)\,, 
\end{equation}
and the resulting top polarization is 
\begin{equation}
   \Pt = \frac{((h_t\,\cos\theta_{\sbottom})^2-(h_b\,\sin\theta_{\sbottom})^2)\,f_1}
			 {\left[(h_t\,\cos\theta_{\sbottom})^2+(h_b\,\sin\theta_{\sbottom})^2-h_t\,h_b\,\sin 2\theta_{\sbottom}\,f_2\right]} \,. \label{eq:lim_higgsino}
\end{equation}
Therefore, for a higgsino-like chargino, the top polarization depends on the sbottom mixing. In the limit of pure LH or RH sbottoms we have 
\begin{eqnarray}
   \tilde b_L &:&  \cos\theta_{\sbottom} = 1 \,,\quad  \Pt \rightarrow +f_1 \,,\\
   \tilde b_R &:&  \cos\theta_{\sbottom} = 0 \,,\quad  \Pt \rightarrow -f_1 \,.
\end{eqnarray}

In other words, a wino-like chargino, whose interaction conserves chirality, couples to a left-chiral sbottom and a left-handed top  (recall that just like the $W^{\pm}$, a wino has only left-chiral interactions) thus always giving a top polarization close to $-1$, similar to single top production in the SM. 
The higgsino interaction, on the other hand, flips chirality and couples a right (left) chiral sbottom to a left (right) handed top; the top polarization thus can vary from $-1$ to $+1$ depending on the sbottom mixing angle and the bottom Yukawa coupling, which becomes relevant at large $\tan\beta$. This will be discussed in more detail in Section~\ref{sect:Num_analysis}.

\subsection{\mbf $\stop \to t\nt$ decays} \label{sect:Pf}

For completeness we also summarize the case of stop decays into neutralinos, c.f.~\cite{Godbole:2011vw}. 
Analogous to eq.~\eq{Pfprime}, the average top polarization from $\stop_i\to t\,\tilde\chi^0_n$  
is given by 
\begin{equation}
       \Pt    = \frac{\big[\, {(b_{in}^{\stop})}^2-{(a_{in}^{\,\stop})}^2 \,\big] \,f_1}
                  { {(b_{in}^{\stop})}^2+{(a_{in}^{\,\stop})}^2-2\,b_{in}^{\stop}\,a_{in}^{\stop}\,f_2} 
   \label{eq:Pf}
\end{equation}
with the obvious replacement $p_{\chm_j}\to p_{\nt_n}$ in eq.~\eq{kinegeneral}. 
Using eqs.~\eq{aik}, \eq{bik} and \eq{Rsf},  the stop couplings are written  as 
\begin{eqnarray} 
  a_{1n}^{\ti t} &=&    -\frac{1}{\sqrt{2}} \left(  N_{n2}+\frac{1}{3} \tan\theta_W  N_{n1}\right)\cos\t_{\tilde t}   - h_t N_{n4}  \sin\t_{\tilde t}\nonumber\\
  b_{1n}^{\ti t} &=&    \frac{2\sqrt{2}}{3}  \tan\theta_W  N_{n1} \sin\t_{\tilde t}   - h_t N_{n4}^*  \cos\t_{\tilde t}
\end{eqnarray}
Substituting these in the expression for the polarization, eq.~\eq{Pf}, one easily sees that in the case 
of large mass differences between the stop and the neutralino ($f_1\rightarrow 1$)
a RH stop  will lead to $\Pt = -1$ when it decays into  a
 higgsino and to $\Pt = +1$ when the decay is into a gaugino; 
a LH stop will lead to the opposite polarizations. 

\clearpage
\section{\mbf Effect of top polarization on the decay kinematics} \label{sect:observables}

The $V-A$ interaction involved in the decay of the top quark, $t\rightarrow Wb\rightarrow i\, i'\, b$, 
where $i$ and $i'$ denote the decay products of the $W$, implies definite correlations between the 
direction of top spin and the top decay products. 
These  are most clearly  understood  in the rest frame of the top quark. Since the top decays before it hadronises, these correlations are not washed out  by the hadronisation process. We discuss explicitly  only the 
case of the top, given that top and anti-top can be distinguished by the charge of the decay lepton. 
Note also that we neglect  the effect of the off-diagonal elements of the CKM matrix
and use $BR(t \rightarrow Wb) = 1$.  

Consider a top quark ensemble with degree of polarization $\P_{t}$.
In the top quark rest frame,  the angular distribution of the decay product $f$  is given by:
\begin{equation}
\frac{1}{\Gamma_t}\frac{\mathrm{d}\Gamma_l}{\mathrm{d}\cos\theta_{f,\rm rest}}=\frac{1}{2}\left(1+ \kappa_f \P_t\cos\theta_{f,\rm rest}\right),\label{eq:topdecay}
\end{equation}
where $\Gamma_t$ is the partial decay width of the top 
and $\theta_{f,\rm rest}$ denotes the angle between the momentum of the 
$f$ and the top spin vector. $\kappa_f$ is called  analyzing power of the decay product $f$~\cite{Bernreuther:2008ju}. 
It is $1$ for a positively charged lepton or a $d$ quark. For a $u$ quark or $\nu_l$,  $\kappa_{u,\nu}=-0.31$, 
while for $b$ and $W$ the values are $\kappa_b=-0.4$ and $\kappa_W=0.4$, respectively. 
The maximal value of $\kappa_{l}=1$ means that the charged lepton is the most efficient polarisation analyzer. 
Corrections to the values of $\kappa_f$ can originate from any deviation of the  $tbW$ coupling from the 
standard $V-A$ structure  and/or from higher order QCD and QED corrections. The leading  QCD corrections 
to $\kappa_b$, $\kappa_d$ and $\kappa_u$ are of the order of a few percent, somewhat decreasing its magnitude~\cite{Brandenburg:2002xr}. 
The values of $\kappa_l$ and $\kappa_{d}$ on the other hand do not receive any corrections from the anomalous $tbW$ coupling at leading order~\cite{Godbole:2010kr}.  Hence the angular distribution of the decay lepton in the rest frame reflects  the polarization of the decaying top quark faithfully, even in the presence of such anomalous couplings. Thus it is an unambiguous measure of top polarization effects. 

It is also interesting to consider how polarization affects the kinematic distributions of the top decay products in the laboratory frame.  This allows to construct useful polarization-dependent observables. The  use of laboratory observables to measure top polarization would obviate the need for reconstruction of the top rest frame.  This is desirable as such a reconstruction  may not be always possible.
The correlation between the polarization of the top and the different  kinematic variables of the decay products can be obtained from  eq.~(\ref{eq:topdecay}) and appropriate Lorentz transformations.  Just like the angular distribution in the top rest frame, the energy integrated decay lepton angular distributions in the laboratory  frame are also unaltered to  linear order in the anomalous $tbW$ 
coupling~\cite{Grzadkowski:2001tq,Grzadkowski:2002gt,Godbole:2002qu,Godbole:2006tq,Godbole:2010kr}.
For all the other distributions, including the energy distribution of the decay lepton, a deviation of the distribution from the unpolarized case can not be uniquely attributed to the top polarization.

While constructing polarization dependent observables, it is worth recalling that  the decay product  distributions 
in the lab frame are influenced not only by the top quark  polarization, but also by the boost $\beta_t$ 
from the top-quark rest frame to the laboratory frame and by the transverse momentum $p_t^T$ of the top quark. 
Here we will use a boost parameter based on the total momentum of the top $|\vec{p}_t|$ and the top energy $E_t$
\begin{equation}
   \beta_t=\frac{|\vec{p}_t|}{E_t}.
\label{Bdef}
\end{equation}

As an example we consider the lab-frame polar angle $\theta_l$ of the lepton w.r.t.\ the top quark direction. Due to the top boost, $\theta_l$ is smaller than its counterpart in the rest frame $\theta_{l,\rm rest}$. Thus, the distribution of 
$\theta_l$ in the lab frame is more strongly peaked towards $0$ for a stronger  top boost as well as for a more positively polarized top quark. One can then define a polar angle asymmetry $A_{\theta_l}$ as

\begin{equation}
   A_{\theta_l}=\frac{\sigma(\theta_l<\pi/4)-\sigma(\theta_l>\pi/4)}{\sigma(\theta_l<\pi/4)+\sigma(\theta_l>\pi/4)} \,.
\label{athetadef}
\end{equation}

In addition to the polar angle, one can study the azimuthal angle distribution. 
To this end, we choose the proton beam direction as the $z$-axis and define 
the top production plane as the $x-z$ plane. Moreover, we identify 
the positive $x$ component with the direction of the top quark.
At the LHC, since the initial state has identical particles, 
the $z$-axis can point in the direction of either proton. This symmetry
implies that one cannot distinguish between an azimuthal angle $\phi$ and
an angle $2\pi- \phi$. 
 In the rest frame this variable does not depend on the 
longitudinal polarization, but in the lab frame it picks up a dependence on 
$\theta_{l,\rm rest}$ through the top boost.  For positively polarized tops it is peaked at $\phi_l=0$ and $\phi_l=2\pi$, with a minimum at $\phi_l=\pi$~\cite{Godbole:2006tq,Godbole:2010kr}. 
It should be noted that 
nonzero $p_t^T$ also causes the $\phi_l$ distributions to peak near $\phi_l=0$
and $\phi_l = 2\pi$, {\it independent} of the polarization state of the
$t$ quark. In other words, the peaking at $\phi_l=0$ and $2 \pi$ is caused by kinematic effects, even for an unpolarized top. It is enhanced even further for a positively polarized top. For a completely negatively polarized top, the pure polarization dependent effects can sometimes even overcome the peaking caused by kinematical effects. The peaks
of the distribution then shift a little away from $\phi=0$ and $2 \pi$. More importantly they lie below those expected for the positively polarized and unpolarized top. The relative number of leptons near $\phi=0$ and $2 \pi$ is thus reduced progressively as we go from a positively polarized to unpolarized to a negatively polarized top. For normalized distributions the ordering is exactly the
opposite at $\phi =\pi$ where the relative number of leptons increases as we go from a positively polarized top to a negatively polarized top.
This shape then motivates the 
definition of the azimuthal angle asymmetry~\cite{Godbole:2010kr}:
\begin{equation}
  A_{\phi_l}=\frac{\sigma(\cos\phi_l>0)-\sigma(\cos\phi_l<0)}{\sigma(\cos\phi_l>0) +\sigma(\cos\phi_l<0)},
\label{Aldef}
\end{equation}
where $\sigma$ is the fully integrated cross section. Note that a higher top polarization or a stronger top boost will result in a more sharply peaked $\phi_l$ distribution and thus yield a  larger asymmetry.

It is also useful to consider energy observables. Although they are 
not completely independent of an anomalous $tbW$ coupling as mentioned above,  
they do carry information about the top polarization. In fact, for a positively polarized top, the energy and the transverse momentum distributions for the lepton are shifted to higher values as compared to the unpolarized or negatively polarized case. 
Since top quarks produced from SM processes are either unpolarized or negatively polarized, this 
feature of the $E_{l}$ and $p_{T}^{l}$ distributions for positively polarized top quarks  can provide an 
effective discrimination against the SM background.
Since the $\kappa_{b}$ and $\kappa_{l}$ have opposite signs, the effect of top polarization on the energy and $p_{T}$ distributions of the $b$--jet in the laboratory frame is exactly in the opposite direction to that for the lepton distributions. In Ref. \cite{Almeida:2008tp}, this feature was employed in constructing  
a discriminator of top quark polarization using the $p_{T}$ of the 
$b$--quark.
Furthermore, the energy distribution can be of particular use when the top quarks are highly boosted. 
In this case, the effect of the boost on the angular distribution may mask the polarization and an accurate 
determination of the angles (for asymmetries) may be difficult. 
It was shown in~\cite{Shelton:2008nq} that in a kinematic regime where the 
tops are heavily boosted, the ratios 
\begin{equation}
   z=\frac{E_b}{E_t},\quad u=\frac{E_l}{E_l+E_b},
\label{zudef}
\end{equation}
are sensitive to the polarization state of the top quark. 
Here $E_t$, $E_b$ and $E_l$ are respectively the (lab frame) energies of the 
top quark, and of the $b$ quark and lepton coming from its decay. The analysis
of~\cite{Shelton:2008nq} was at the LO parton level, but in practical 
applications one may also consider $E_b$ to be the energy of {\it e.g.}\ a $b$ jet. 
Note that the ranges of $z$ and $u$ are given in principle by
\begin{equation}
   0\leq z,u\leq 1,
\label{zurange}
\end{equation}
although there will be a cut-off at high and low values due to the finite $b$ quark and $W$ boson masses.  In the collinear limit $\beta_{t} = 1$, the normalized distribution  ${\frac{1}{\Gamma}} {\frac{d\Gamma}{dz}}$ can be in fact computed analytically~\cite{Shelton:2008nq}. It  
is peaked  at lower  values of $z$ for a positively polarized  tops, and at high values of $z$ for  negatively polarized tops. In case of the $u$ distribution, which
has to be computed numerically, even in the  $\beta_{t} =1$ limit,  the peak is shifted by about $0.1$ for $\P_{t} = -1$ compared to the unpolarized case; whereas for $\P_{t} = 1$ the normalized distribution is weighted towards 
larger values of $u$.  One can of course  define these observables for
any value of a cut on the top boost. However, at low boost values, both $z$ and $u$ are
increasingly contaminated with contributions that are independent of  $\Pt$, thus reducing their effectiveness as discriminators of top polarization and/or new physics parameters. 
We will show later how these distributions may be exploited for quantitative measures of polarization.

\clearpage
\section{\mbf Boost treatment} \label{sect:boost_treatement}

Before proceeding to the numerical analysis, we find it useful to give some details on the treatment of the boost. 
Note that the overall boost of the top quark in the laboratory frame depends on the boost of the sfermion, that of the top in the sfermion rest frame and the angle of emission of the top with respect to the sfermion. As a consequence, the relation between the top polarization in the sfermion rest frame, discussed in Sections~\ref{sect:Pfsfrest}  and \ref{sect:sbottom-num}, and the top polarization measured in the lab frame depends on all these. 

We explain the procedure we have used to obtain the top quark polarization in the lab frame with the concrete example of $\sbottom_1\to t \chm_1$ decays. 
The same treatment will later be applied to all other decays, including stop decays to top+neutralino, in order 
to obtain the total polarization from all sbottom and stop decays into tops, \ie\ the quantity that is actually 
relevant for experiment.

First, we define our frame of reference such that the decaying sbottom is at rest and the top momentum lies in the $x-z$ plane. In this frame, the momentum vectors of the sbottom, top and chargino are defined as: 
\begin{eqnarray}
p_{\tilde b_1} &=& (m_{\tilde b_1}, 0\,, 0\,, 0) \\
p_t &=& (E_t,\, |p_t|\sin\vartheta\,,0\,,|p_t|\cos\vartheta) = (E_t, \, p_{t_x},0\,,p_{t_z}) \nonumber\\ 
p_{\tilde\chi_1^-} &=& (E_{\tilde\chi_1^-}, -|p_t|\sin\vartheta\,,0\,,-|p_t|\cos\vartheta) \nonumber
\end{eqnarray}
where, $\vartheta$ is the angle between decaying top and chargino and $|p_t|$ is the magnitude of the top momentum given by: 
\begin{equation}
|p_t| = \frac{\lambda^{\frac{1}{2}}(\msb1^2,m_t^2,m_{\ch_1}^2)}{2\,\msb1}
\end{equation} 
Furthermore, we specify the $z$-component of the top spin vector as:\footnote{Since we are interested in the longitudinal polarization of the top, we are not concerned about the spin vectors in the $x$ and $y$ directions.}
\begin{equation}
s^3_t = \left( \frac{|p_t|}{m_t},\,\frac{E_t}{m_t}\,\frac{p_{t_x}}{|p_t|},0,\frac{E_t}{m_t}\frac{p_{t_z}}{|p_t|}\right)
\end{equation} 
\noi
The top-polarization in terms of helicity amplitude formalism is defined as:
\begin{eqnarray}
\Pt &=& \displaystyle\frac{\int\frac{d\sigma(+,+)}{d\cos\vartheta}\,d\cos\vartheta - \int\frac{d\sigma(-,-)}{d\cos\vartheta}\,d\cos\vartheta}{\int\frac{d\sigma(+,+)}{d\cos\vartheta}\,d\cos\vartheta + \int\frac{d\sigma(-,-)}{d\cos\vartheta}\,d\cos\vartheta} \nonumber\\
& = & \displaystyle\frac{(k_{ij}^2-l_{ij}^2)\,\int m_t\,(p_{\tilde\chi_1^-}\cdot\, s^3_t)d\cos\vartheta}{(k_{ij}^2+l_{ij}^2)\,\int (p_t\cdot\,p_{\tilde\chi_1^-})d\cos\vartheta-2k_{ij}\,l_{ij}\int m_t\,m_{\tilde\chi_1^-}d\cos\vartheta}
\end{eqnarray}
Now we boost the system in the positive $z$-direction. Note that the spin vector $s^3_t$ is not a Lorentz vector, hence the dot product $(p_{\tilde\chi^-_1}\cdot s^3_t)$ is not Lorentz invariant. Thus, the resulting top polarization has a dependence on the boost.

\clearpage
\section{\mbf Numerical analysis} \label{sect:Num_analysis}

For the numerical analysis, we choose to work within the framework of the general R-parity and CP conserving 
MSSM with parameters defined at the electroweak scale.  
The relevant soft terms for our analysis are the left and right 3rd generation soft masses
$M_{\ti Q_3}$, $M_{\ti U_3}$, $M_{\ti D_3}$ and the trilinear couplings $A_t$, $A_b$ 
entering the stop and sbottom mass matrices, 
together with the gaugino masses $M_1$ and $M_2$, the higgsino mass parameter $\mu$, and $\tan\beta$. 
The top quark mass is fixed at $173.2$~GeV. 

We consider the two cases of higgsino-like or wino-like $\tilde\chi^\pm_1$.
For the example of a higgsino-like chargino, we set  $\mu = 350$~GeV and $M_2 = 1000$~GeV; the neutralino sector is fixed by $M_1 = 500$~GeV.  This gives $m_{\tilde\chi^\pm_1}\simeq352$~GeV, $m_{\tilde\chi^0_2}\simeq358$~GeV and
$m_{\tilde\chi^0_1}\simeq 343$~GeV (for $\tan\beta=10$, but showing only little variation with $\tan\beta$). 
For the wino-like case, we reverse the parameters, setting $M_2 = 350$~GeV and $\mu = 1000$~GeV. 
With $M_1 = 250$~GeV, this gives $m_{\tilde\chi^\pm_1}\simeq m_{\tilde\chi^0_2}\simeq360$~GeV and $m_{\tilde\chi^0_1}\simeq 247$~GeV. 
\footnote{When the rest of the SUSY spectrum is needed, we set the slepton soft terms to 
$M_{\ti L_i}=M_{\ti R_i}=500$~GeV ($i=1...3$),  the squark soft terms of the first two generations to 
$M_{\ti Q_j}=M_{\ti U_j}=M_{\ti D_j}=2$~TeV ($j=1,2$), the gluino soft mass $M_3=1.5$~TeV, and the pseudoscalar Higgs mass $m_A=1.5$~TeV. The right stop mass parameter $M_{\ti U_3}$ and the trilinear coupling $A_t$ are adjusted such that $m_h\approx 126$~GeV.}

In general, we use {\tt SoftSUSY}~\cite{Allanach:2001kg} to compute the full MSSM spectrum.  
However, when considering only the $\sb_1\to\chm_1t$ decay, we directly take the lighter sbottom mass, $\msb{1}$, and the sbottom mixing angle, $\cos\theta_{\sb}$, as free parameters; the exact values of the stop/sbottom soft terms are not necessary in this case. For computing sparticle decays, checking flavor observables, etc., 
we use {\tt micrOMEGAs}~\cite{Belanger:2004yn,Belanger:2010gh}. 
Cross sections are computed at NLO with {\tt Prospino}~\cite{Beenakker:1996ed}. 
Finally, for generating distributions, we use {\tt MadGraph}~\cite{Alwall:2007st,Alwall:2011uj} with v4 model files. 

\subsection{\mbf Parameter dependencies for $\sb_1$} \label{sect:sbottom-num}

\begin{figure}[t]\centering
  \includegraphics[width=0.5\textwidth]{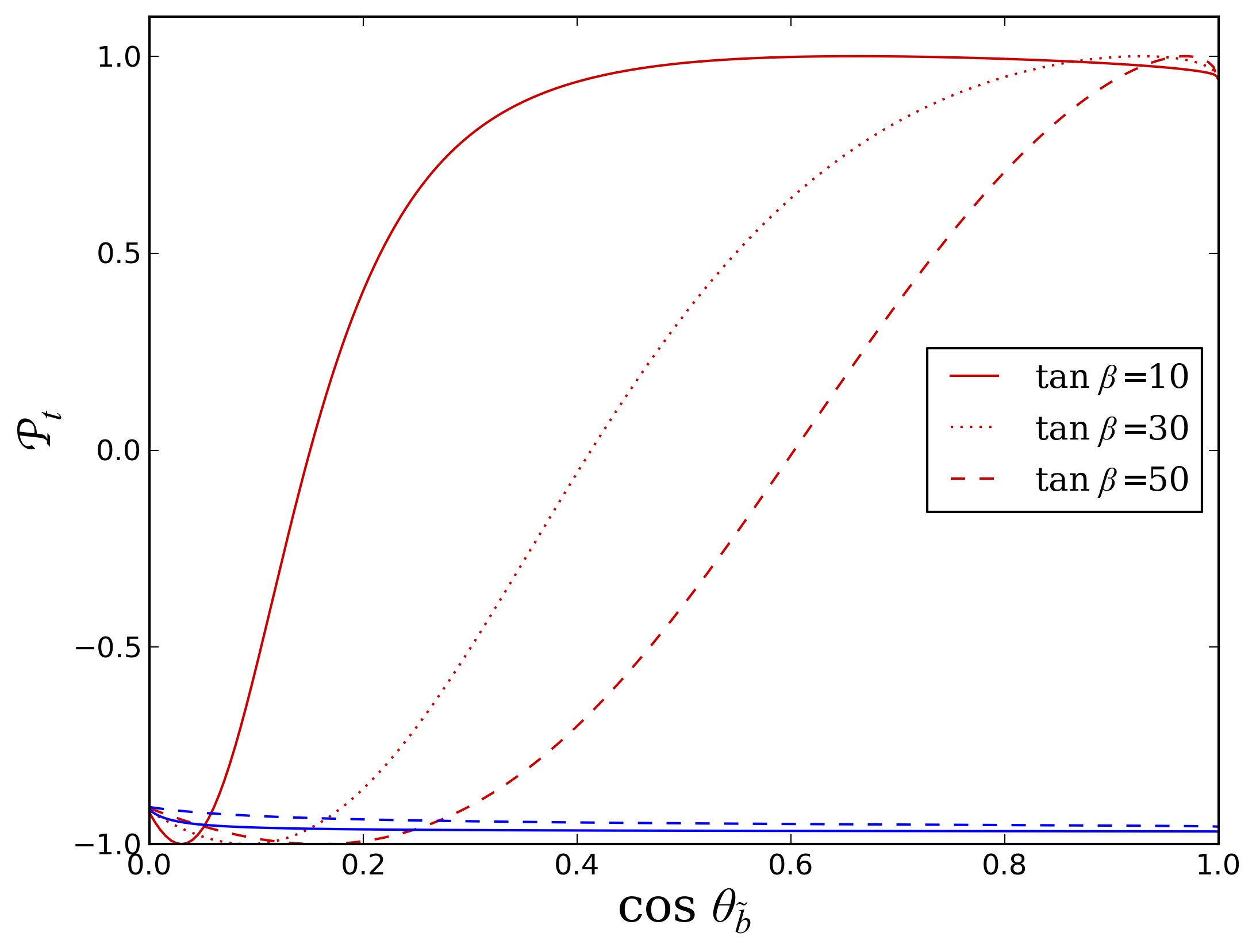}  
\caption{Top polarization in $\sb_1\to\chm_1t$ decays as a function of the sbottom mixing angle, for $\msb{1}=685$~GeV. 
For the higgsino-like case $\mu = 350$~GeV and $M_2 = 1000$~GeV (in red), while for the wino-like case 
$M_2 = 350$~GeV and $\mu = 1000$~GeV (in blue). 
The solid, dotted, and dashed lines are for $\tan\beta = 10$, $30$, and $50$, respectively. 
\label{fig:mixing_pol}}
\end{figure}

Let us first discuss the results for top polarization in the sbottom rest frame. 
Figure~\ref{fig:mixing_pol} shows the polarization of the top quark coming from the $\sb_1\to\chm_1t$ decay as a function of the sbottom mixing angle, for the case of large $\Delta m$; concretely we take $\msb{1}=685$~GeV and, as mentioned above, $\mu = 350$~GeV and $M_2 = 1000$~GeV ($M_2 = 350$~GeV and $\mu = 1000$~GeV) for the higgsino (wino) case, with $M_1 = 500$~GeV. As discussed in Section~\ref{sect:Pfpr}, for a wino-like chargino the polarization of the top is always $\approx -1$ regardless of the nature of the decaying sbottom. For a higgsino-like chargino, on the other hand, $\Pt$ varies from $-1$ for a $\sb_L$ to $+1$ for a $\sb_R$. For small $\tan\beta$, i.e.\ small bottom Yukawa coupling, the transition happens very quickly around $\cos\theta_{\sb}\sim0.1-0.25$. For large $\tan\beta$, where top and bottom Yukawa couplings are equally important, the transition happens more  slowly and for maximally mixed sbottoms; 
as a result, all values of $\Pt \in [-1,+1]$ can be easily obtained without having to fine-tune the value of the sbottom mixing angle.  We also note that for $\cos\theta_{\sb}=0$, $\Pt=-0.92$; when $\cos\theta_{\sb}$ slightly increases above zero, the top Yukawa coupling starts playing a role in the denominator of eq.~(\ref{eq:lim_higgsino}), causing $\Pt$ to first decrease to $-1$ before increasing to $+1$ for larger $\cos\theta_{\sb}$. 
A similar effect can be seen for $\cos\theta_{\sb}\approx 1$:  $\Pt=+1$ is obtained for $\cos\theta_{\sb}=0.7$, while at 
$\cos\theta_{\sb}=1$, $\Pt=0.945$. 

\begin{figure}[h]\centering
  \includegraphics[width=0.5\textwidth]{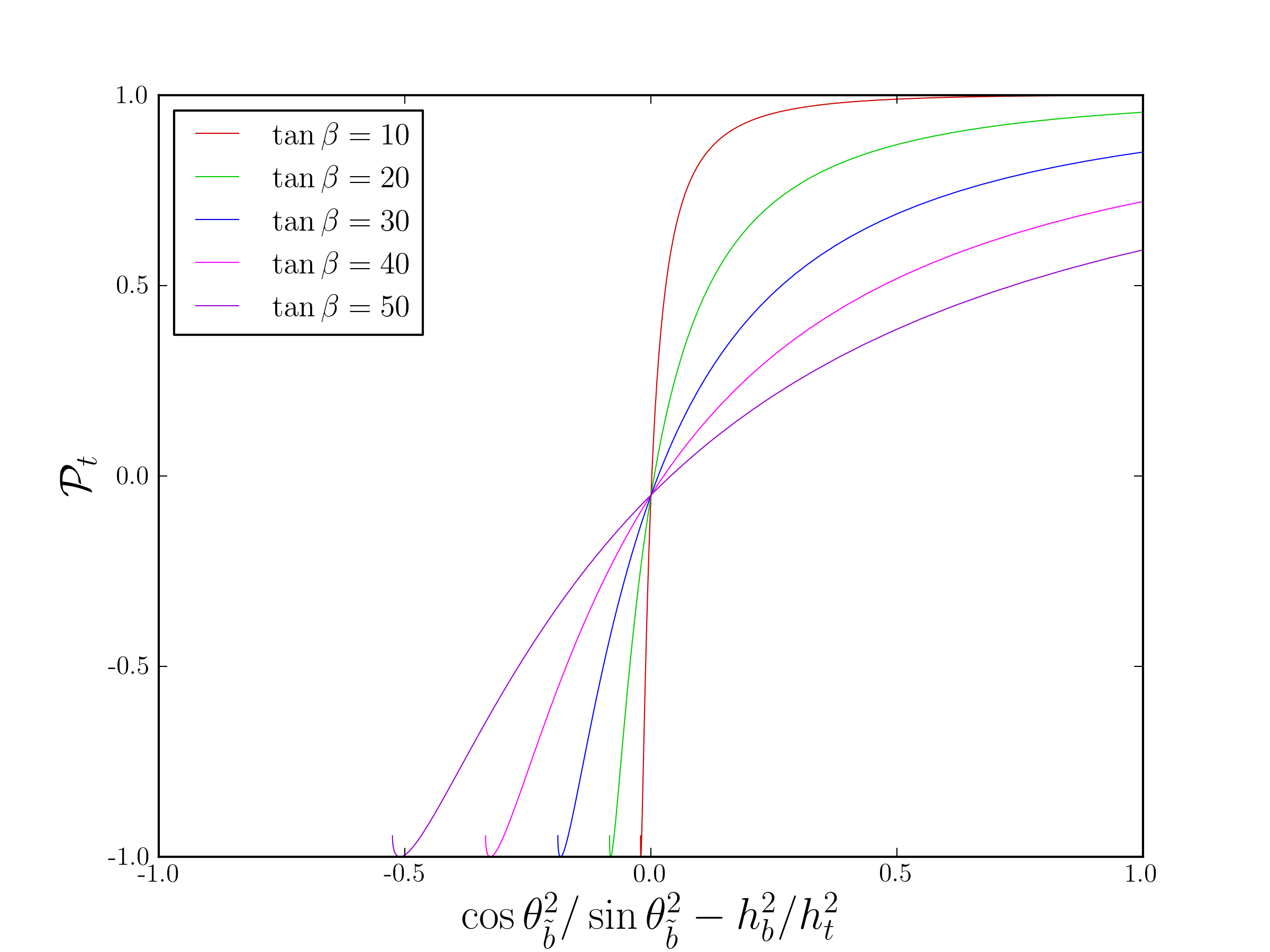}  
\caption{Dependence of the top polarization in $\sb_1\to\chm_1t$ decays on the relative strength of the $t$ and $b$ Yukawa couplings and the sbottom mixing angle. 
\label{fig:yuk_mixing}}
\end{figure}

In order to understand the dependence of $\Pt$ on the relative strength of the Yukawa couplings, let us simplify eq.~(\ref{eq:Pfprime}) for a completely higgsino-like chargino, cf.\  eq.~(\ref{eq:lim_higgsino}).   
We get 
\begin{equation}
   \Pt = \frac{\left(\frac{\cos^2\theta_{\sbottom}}{\sin^2\theta_{\sbottom}}-\frac{h_b^2}{h_t^2}\right)\,f_1}
                      {\left(\frac{\cos^2\theta_{\sbottom}}{\sin^2\theta_{\sbottom}}+\frac{h_b^2}{h_t^2}\right)-2\,\frac{\cos\theta_{\sbottom}}{\sin\theta_{\sbottom}}\frac{h_b}{h_t}\,f_2} 
\end{equation}
Now recall that $f_1 \simeq 1$ away from the kinematic boundary, while $f_2\to 0$.
Hence, for $\cos^2\theta_{\sbottom}/\sin^2\theta_{\sbottom} > h_b^2/h_t^2$, $\Pt > 0$ and for $\cos^2\theta_{\sbottom}/\sin^2\theta_{\sbottom} < h_b^2/h_t^2$, $\Pt < 0$.  This effect is illustrated in Fig.~\ref{fig:yuk_mixing}. Also, for a fixed value of $\cos\theta_{\sbottom}$, $|\Pt|$ decreases with increasing $\tan\beta$.

\begin{figure}[t!]\centering
   \includegraphics[width=0.49\textwidth]{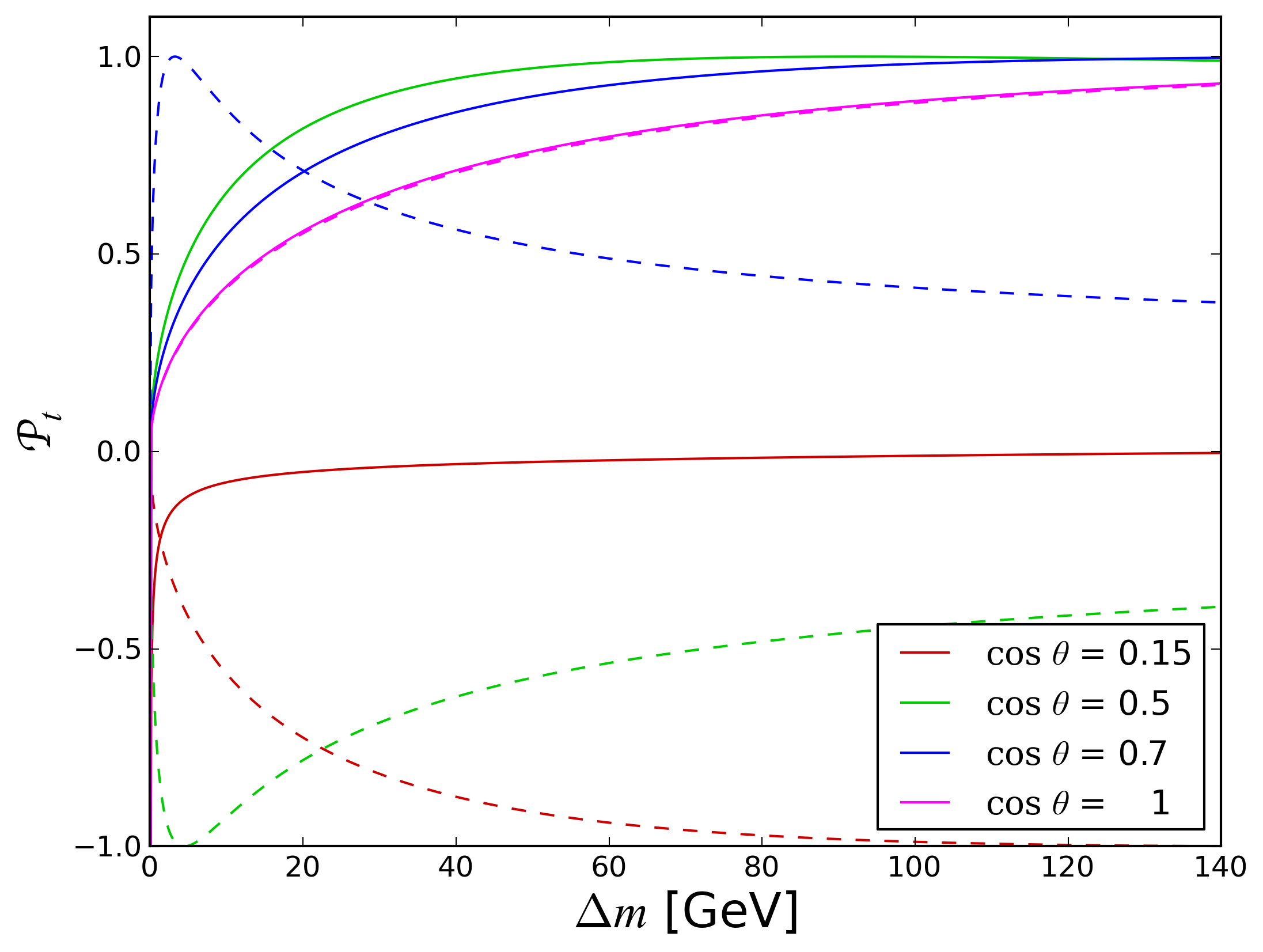}\hspace*{-1mm}\includegraphics[width=0.49\textwidth]{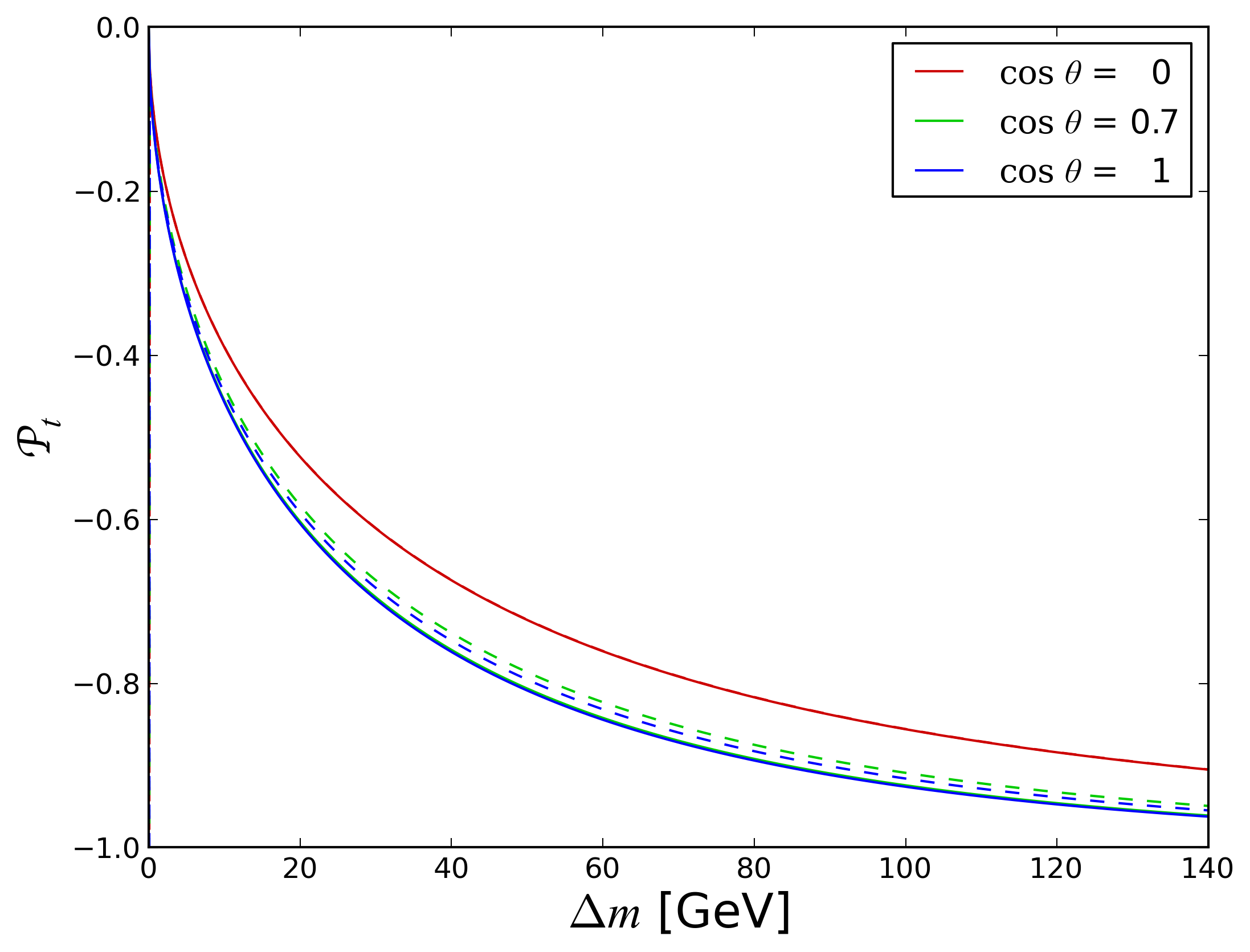}
\caption{Top polarization in $\sb_1\to\chm_1t$ decays as a function of $\Delta m\equiv m_{\sbottom_1}-m_{\chm_1}-m_{t}$ for various choices of  $\cos\theta_{\sb}$; on the left for $\mu = 350$~GeV and $M_2 = 1000$~GeV (higgsino case), on the right for $M_2 = 350$~GeV and $\mu = 1000$~GeV (wino case). Solid lines represent $\tan\beta = 10$ and dashed lines are for $\tan\beta = 50$. \label{fig:kinemtics_pol}}
\end{figure}
\begin{figure}[t!]\centering \vspace*{3mm}
\hspace*{-2mm}\includegraphics[width=0.5\textwidth]{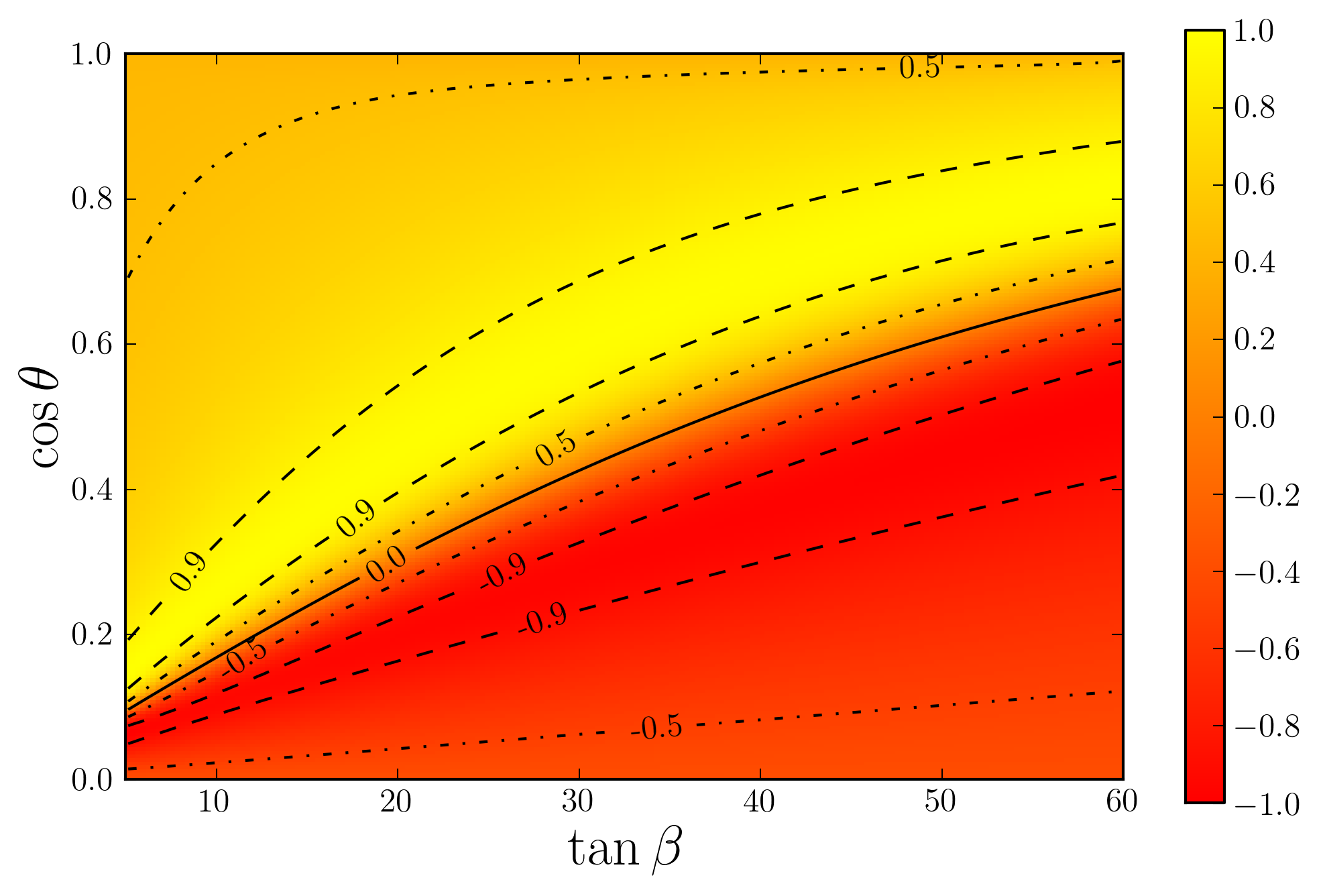}\hspace*{-1mm}\includegraphics[width=0.5\textwidth]{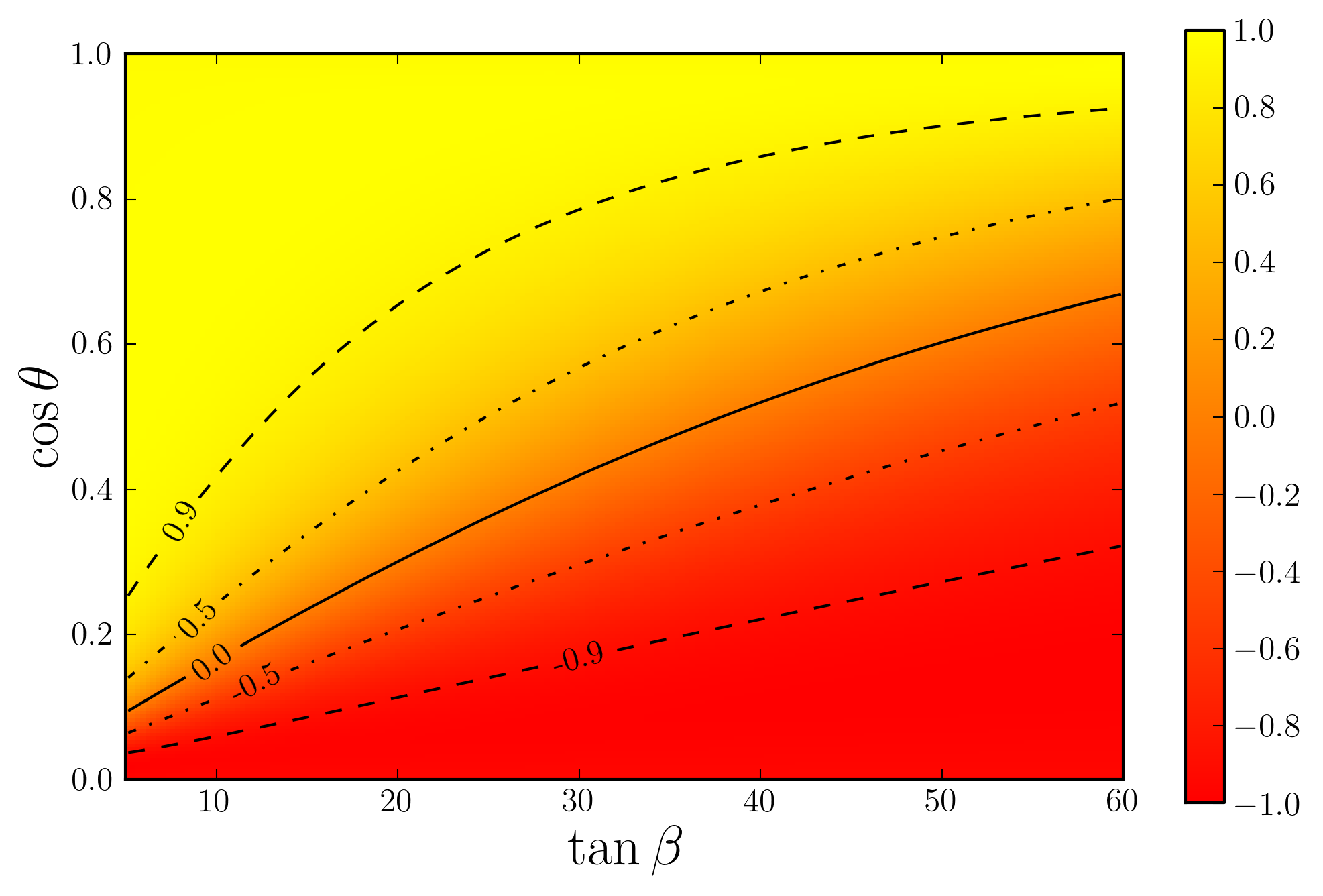}
\caption{Contours of top polarization in the $\cos\theta$ versus $\tan\beta$  plane for the higgsino case with $\mu = 350$~GeV and $M_2 = 1000$~GeV, for a fixed mass difference of $\Delta m=10$~GeV (left) and $\Delta m=200$~GeV  (right). \label{fig:contours_2}}
\end{figure}

The dependence on the available phase space, $\Delta m$, is illustrated in Fig.~\ref{fig:kinemtics_pol}.
The left panel shows the case of a higgsino-like chargino while the right panel shows the wino-like case. The top polarization as described in eq.~(\ref{eq:Pfprime}) is directly proportional to $f_1$, which goes to zero when $\Delta m \rightarrow 0$ and, as a result, $\Pt \rightarrow 0$. This can very well be seen in both the higgsino-like and the wino-like scenarios. Away from the kinematic boundary, $f_1 \rightarrow 1$ and $f_2 \rightarrow 0$. For $\Delta m > 100$~GeV, $f_1$ is very near 1 and  the polarization approaches its limiting value. Here, the contribution of
 the  $f_2$  eq.~(\ref{eq:lim_higgsino}) is somewhat suppressed by the
 ratio of Yukawa couplings or the sbottom mixing angle.
Note also that in the higgsino case, for certain values of the mixing angle, {\it e.g.}\ $\cos\theta_{\sbottom}=0.5$,  increasing the value of $\tan\beta$ can flip the sign of the polarization, this corresponds to the transition  $\cos^2\theta_{\sbottom}/\sin^2\theta_{\sbottom} =h_b^2/h_t^2$.
Contours for the top polarization in the $\cos\theta_{\sbottom}$  vs $\tan\beta$ plane are displayed in Fig.~\ref{fig:contours_2} for both small and large values of $\Delta m$. In both cases any value of $P_t$ can be reached for any value of $\tan\beta$, although for $\Delta m=10$~GeV and small $\tan\beta$ the polarization changes very rapidly with the sbottom mixing angle as a result of the kinematic factor $f_2$ in the denominator of eq.~(\ref{eq:lim_higgsino}).

\begin{figure}[t]\centering
  \hspace*{-2mm}
  \includegraphics[width=0.48\textwidth]{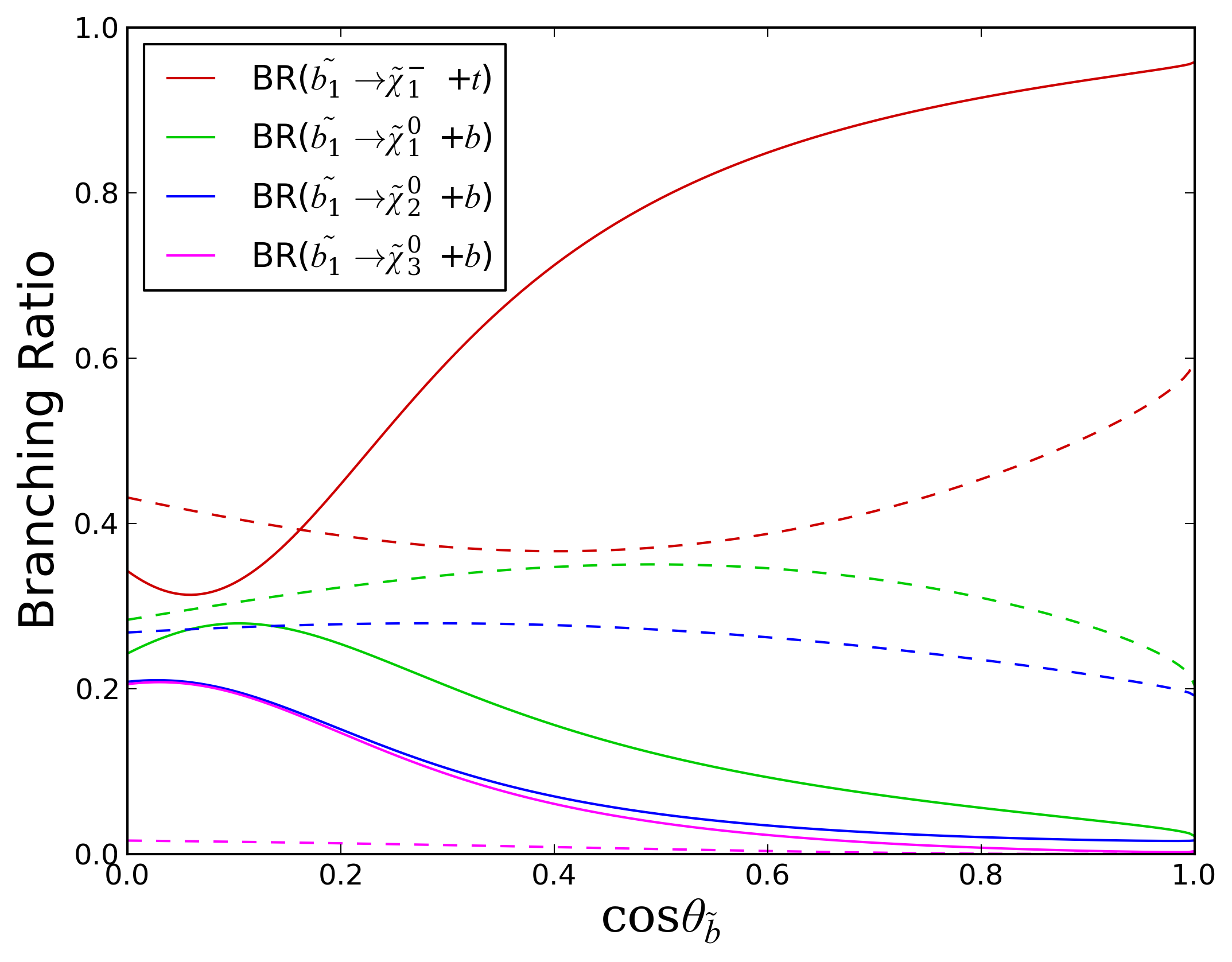}
  \includegraphics[width=0.48\textwidth]{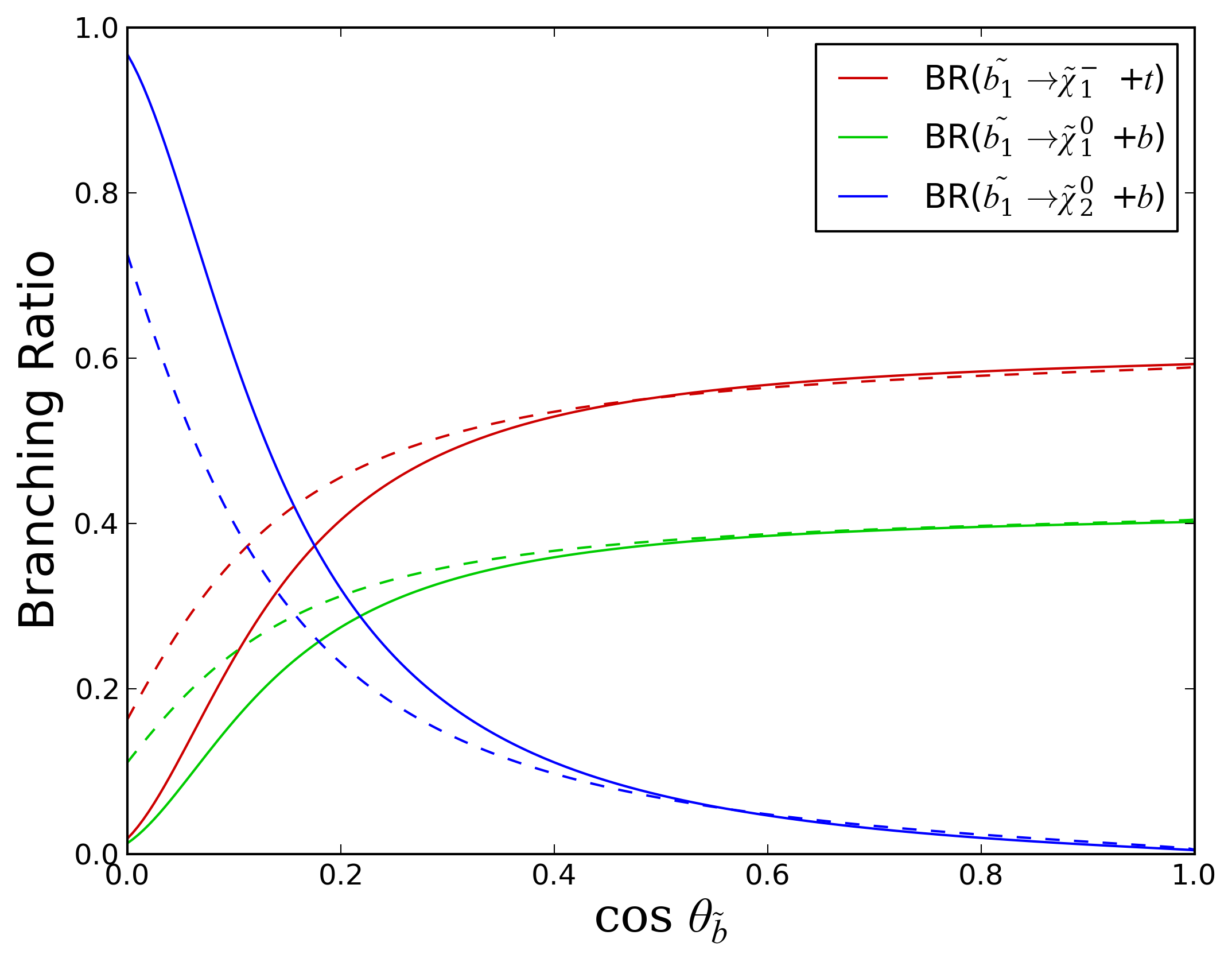}  
\caption{Branching ratios of $\sb_1$ as a function of $\cos\theta_{\sb}$ for $\msb{1}=685$~GeV,  
on the left for the higgsino case with $\mu = 350$~GeV and $M_2 = 1000$~GeV, 
on the right for the wino case with $\mu = 1000$~GeV and $M_2 = 350$~GeV. 
The solid (dashed) lines represent $\tan\beta = 10$ ($50$). 
The remaining parameters are as for benchmark point {\tt bm-2} in Table~\ref{tab:bm}.
\label{fig:mixing_br}}
\end{figure}

To see whether there is any hope to actually measure this polarization, we also need to consider the sbottom production cross section and decay branching ratios. The branching ratios for $\sb_1$ decaying into charginos and neutralinos as function of the sbottom mixing angle are depicted in  Fig.~\ref{fig:mixing_br}. The plot on the left shows the higgsino-like case. At small $\tan\beta$, the decay into $\tilde \chi^-_1t$ is always dominant even though the branching ratio is reduced when $\sb_1\approx \sb_R$ ($\cos\theta_{\sb} \approx 0$), because the coupling strength of a $\sb_R$ to a higgsino is proportional to the bottom Yukawa coupling. Therefore, at small $\cos\theta_{\sb}$, sbottom decays into charginos and each of the three lightest neutralinos become comparable (about 20--30\%). 
For $\tan\beta = 50$, the bottom Yukawa coupling is large and the branching ratio to $\tilde \chi^-_1t$ is about 40--60\% over the whole $\cos\theta_{\sb}$ range. \\
In the right panel of Fig.~\ref{fig:mixing_br} the  $\tilde \chi^-_1$ and $\tilde \chi^0_1$ are wino-like, $\tilde \chi^0_2$ is bino-like and $\tilde \chi^0_3, \tilde \chi^0_4$ are higgsino-like. The wino states do not couple to $\sb_R$, so for $\cos\theta_{\sb}\approx 0$ the decay $\sb_1\to\nt_2b$ dominates. As the $\sb_L$ admixture increases, the branching ratio into the wino-like chargino increases because the coupling is proportional to top Yukawa coupling.  The BR$(\sb_L\to\chm_1t)$
saturates at around 60\% because it has to compete with the $\sb_L\to \nt_1b$ decay, which has a BR around  40\%.

\begin{figure}[t]\centering
  \includegraphics[width=0.55\textwidth]{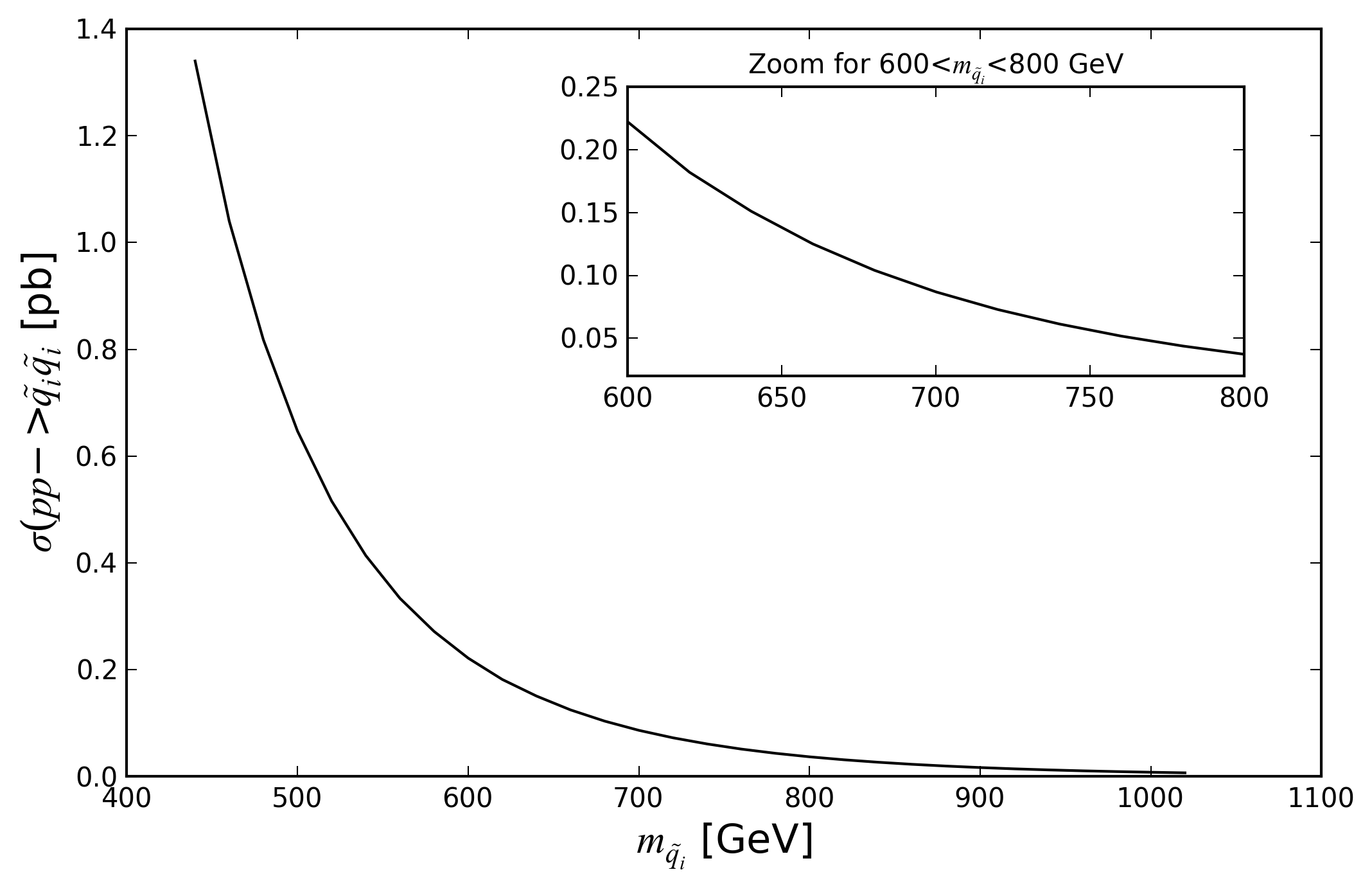}  
\caption{Stop and sbottom pair production cross sections at the LHC with $\sqrt{s}=14$~TeV, 
computed at NLO  with {\tt Prospino}. 
\label{fig:produces}}
\end{figure}

The cross sections for  sbottom (and stop) pair-production as a function of the sbottom (stop) mass, 
calculated at NLO  at $\sqrt s = 14$ TeV with {\tt Prospino}~\cite{Beenakker:1996ed}, are shown in Fig.~\ref{fig:produces}. 
The processes $pp\to \sb_i^{}\sb_i^*$, and likewise $\stop_i^{}\stop_i^*$, ($i=1,2$) proceed through 
pure QCD interaction, the mixing angle in the stop or the sbottom sector enters the cross section calculations only through $\mathcal{O}(\alpha_s)$ corrections involving $t\tilde t \tilde g$ and the four-squark couplings. Thus the cross section for  stop-pair production is very similar to that of sbottom-pair production. Mixed pairs $\tilde t_1 \tilde t_2$ and $\tilde b_1 \tilde b_2$ cannot be produced at lowest order since the $g \tilde t \tilde t$ and $g g \tilde t \tilde t$ vertices are diagonal in the chiral as well as in the mass basis. $\tilde t_i \sbottom_i$ production cross sections are suppressed by at least 2 orders of magnitude as compared to  $\tilde t_i \tilde t_i$ and $\sbottom_i \sbottom_i$ productions, we ignore this production mode for stop and sbottom. The production cross section changes rapidly in the interesting mass range between 500~GeV and 1~TeV. 
For masses around 650 GeV, as considered in the benchmark scenarios below, we find a cross section of about 150~fb, 
which quickly falls off to below 50~fb for around 800 GeV, see the small inset in Fig.~\ref{fig:produces}. 

\clearpage

\clearpage

\subsection{Benchmark scenarios} \label{sect:bm_scenario}

In a complete MSSM scenario, tops can also come from decays of stops, $\tilde t_{1,2}$, and of the 
heavier sbottom, $\tilde b_2$. Some of these decays can be distinguished by their different signatures, 
for instance $\tilde b_1^{} \to t\, \tilde\chi_1^-\to tW\,\tilde\chi_1^0$ versus 
$\tilde t_1^{} \to t\, \tilde\chi_{2}^0 \to t\,Z \tilde\chi_{1}^0$ or $\tilde t_1^{} \to t\,\tilde\chi_{1}^0$. 
Others, like $\tilde b_1^{}$ and $\tilde b_2^{}$ both decaying to $t\, \tilde\chi_1^-$ 
give identical signatures. 
It is hence necessary to consistently add up the polarizations resulting 
from all processes which cannot be distinguished in the experimental analysis. 
In particular, if the masses of the $\tilde\chi_1^\pm$, $\tilde\chi_2^0$ and $\tilde\chi_1^0$ are close 
to each other, as is typically the case in higgsino or wino LSP scenarios, the $\tilde\chi_1^\pm$ and 
$\tilde\chi_2^0$ decays into the LSP will lead to soft decay 
products which are likely to be missed. In such a case, the processes 
\begin{eqnarray}
  p p &\rightarrow& \tilde b_1^{}\tilde b_1^*, \quad \tilde b_1^{} \to t\, \tilde\chi_1^-   \nonumber\\
  p p &\rightarrow& \tilde b_2^{}\tilde b_2^*, \quad \tilde b_2^{} \to t\, \tilde\chi_1^-   \nonumber\\
  p p &\rightarrow& \tilde t_1^{}\tilde t_1^*, \quad \,\tilde t_1^{} \to t\, \tilde\chi_{1,2}^0   
\label{eq:net_pol}
\end{eqnarray}
may all contribute to the $t\bar t+E_T^{\rm miss}$ signature, and the total or ``net'' top polarization  
relevant for the experimental analysis will be a result of the stop and sbottom production cross sections, 
their decay branching ratios  and the relevant boosts for going from the rest frame to the lab frame. 
(We neglect the $\stop_2$ in eq.~(\ref{eq:net_pol}) because, in order to achieve $m_h\approx 126$~GeV, 
at least one of the stops should be heavy and will thus have 
a very low production cross section.) 

In this context, we remind the reader that the situation of only $\sbottom_1$ being light and all other 
3rd generation squarks heavy only occurs for $\sbottom_1\sim \sbottom_R$. For $\sbottom_1\sim \sbottom_L$, 
also the $\stop_1$ will be close in mass (or even lighter because of L--R mixing), 
because both their masses are governed by the same mass parameter $M_{\tilde Q_3}$. 

For illustration and to allow a complete analysis, we present in Table~\ref{tab:bm} a set of 7 benchmark points 
which exemplify different scenarios of stop/sbottom mass and mixing patterns and resulting top polarizations. 
The production cross sections, branching ratios and top polarizations $\P_t$ originating from different 
decay processes in the respective squark rest frame are listed in Table~\ref{tab:net_pol}. 
Moreover, Table~\ref{tab:net_pol} gives the net polarization $\widehat\P_t$ in the laboratory frame, together with 
the values for the polar angle and azimuthal angle asymmetries defined in eqs.~(\ref{athetadef}) and (\ref{Aldef}),  
summing over all processes that cannot be distinguished from $\tilde b_1^{} \to t\, \tilde\chi_1^-$.\footnote{Of course 
dedicated simulations would be necessary to decide whether or not specific processes can be distinguished. 
This is however beyond the scope of this paper. Instead we generally treat $\sbottom_1\to t\,\tilde\chi^-_1$ and 
$\stop_1\to t\,\tilde\chi_{1,2}^0$ as indistinguishable if the $\tilde\chi_1^\pm - \tilde\chi_1^0$ mass difference is small, 
below about 20~GeV.} 

\begin{table}[t] \centering
\begin{tabular}[h]{ | c | r | r | r | r | r | r | r | }
\hline
  & {\tt bm-1} & {\tt bm-2} & {\tt bm-3} & {\tt bm-4} & {\tt bm-5} & {\tt bm-6} & {\tt bm-7}  \\ 
\hline
$M_{\tilde Q_3}$ & 571 & 1300 & 630 & 610 & 582 & 850 &600  \\
$M_{\tilde D_3}$  & 1500 & 572 & 597 & 599&1500 & 601.5 & 609 \\
$M_{\tilde U_3}$ & 2200 & 1300 & 2200 & 2200 & 2200 &1500& 2200  \\ 
\hline
$M_1$ & 500 & 500 & 500 & 500 & 200 & 500& 250  \\ 
$M_2$ & 1000 & 1000 & 1000 &1000 & 400 &1000& 350  \\
$\mu$ & 350 & 350 & 350 &350 &150 & 350 & 1000  \\ 
\hline
$m_{\tilde b_1}$ & 650.41 & 650.44 & 650.85 & 650.13 &650.08 & 650.17 & 650.71  \\
$m_{\tilde b_2}$ & 1537.39 & 1329.77 & 707.36 & 688.72 & 1538.10 & 884.47&  692.45  \\
$\cos\theta_{\tilde b}$ & 0.999 & 0.006 & 0.10  & 0.15 &0.999 & 0.02  & 0.42 \\ 
\hline
$m_{\tilde t_1}$ & 630.18 & 1236.20  & 687.92& 668.37 &633.84 & 818.99& 667.61  \\
$m_{\tilde t_2}$ & 2210.95 & 1560.58 & 2208.66  & 2209.42& 2212.19 & 1518.19 & 2205.97  \\
$\cos\theta_{\tilde t}$ & 0.995 & 0.85 & 0.995  &0.995 & 0.996  & 0.96 &  0.996 \\ 
\hline
$m_{\tilde \chi_1^-}$ & 351.91  & 352.73 & 352.04 &  351.98 & 144.16  & 352.05& 360.35  \\
$m_{\tilde \chi_2^-}$ & 1013.41 & 1021.05 & 1013.45 & 1013.42 &430.32 & 1016.09& 1008.07  \\
$m_{\tilde \chi_1^0}$ & 343.06  & 343.39  & 343.10 & 343.08&126.52 & 343.13&  247.08  \\
$m_{\tilde \chi_2^0}$ & 357.41 & 358.22 & 357.58 & 357.50 & 159.18  & 357.22& 360.10  \\
$m_{\tilde \chi_3^0}$ &502.99 &  500.89 & 502.32 & 502.45 & 213.20  & 502.61&  1002.10  \\ 
$m_{\tilde \chi_4^0}$ & 1013.95  &  1021.33  &  1014.02  & 1013.98 &  431.95 & 1016.63& 1008.24  \\ 
\hline
$\bsgsub$   & $3.53$ & $3.61$ & $3.51$ & $3.51$ & $3.26$ & $3.78$ & $3.31$  \\
$\bmmsub$ & $3.04$ & $3.04$ & $3.03$ & $3.04$ & $3.04$ & $3.03$ & $3.03$  \\ 
\hline
\end{tabular}
\caption{Parameters and masses (in GeV) for seven illustrative benchmarch points. 
All points have $\tan\beta=10$ and $A_b=100$~GeV, while $A_t\approx3$~TeV is adjusted  
such that $m_h\approx126$~GeV.  
The values for $\bsg$ and $\bmm$ are given in units of $10^{-4}$ and $10^{-9}$, respectively.
\label{tab:bm}}
\end{table}

\begin{table}[t] \centering
\begin{tabular}[h]{ | l | r | r | r | r | r| r | r | }
\hline
& {\tt bm-1} & {\tt bm-2} & {\tt bm-3}  & {\tt bm-4} & {\tt bm-5} & {\tt bm-6} & {\tt bm-7} \\
\hline
$\sigma(pp\to\tilde b_1^{} \tilde b_1^*)$ [pb] & 0.137  & 0.137 & 0.137 & 0.137 & 0.137 & 0.137 & 0.137 \\
$\sigma(pp\to\tilde b_2^{} \tilde b_2^*)$ [pb]  & $<10^{-3}$  & 0.001 & 0.082 & 0.096 & $<10^{-3}$ & 0.019 & 0.093  \\
$\sigma(pp\to \tilde t_1^{} \tilde t_1^*)$ [pb]  & 0.163  & 0.002 & 0.095 & 0.114 & 0.157 & 0.033 & 0.115  \\
\hline
${\rm BR}(\tilde b_1 \to t\,\tilde \chi^-_1)$ & 0.96 &0.34  & 0.31 & 0.35 & 0.72 & 0.32 & 0.41  \\
${\rm BR}(\tilde b_1 \to  t\,\tilde \chi^-_2)$ & -- & -- & --  & -- &0.16 & -- & -- \\
${\rm BR}(\tilde b_2 \to  t\,\tilde \chi^-_1)$ & 0.30 & 0.82  & 0.96 &0.96 & 0.29 & 0.96 &0.58  \\
${\rm BR}(\tilde b_2 \to  t\,\tilde \chi^-_2)$ & $<10^{-3}$ & 0.09 & -- & -- & 0.01 & -- & --  \\
${\rm BR}(\tilde t_1 \to  t\,\tilde \chi^0_1)$ & 0.51 & 0.33 & 0.50 & 0.50 & 0.27 & 0.47 & 0.01  \\
${\rm BR}(\tilde t_1 \to  t\,\tilde \chi^0_2)$ & 0.45 & 0.35 & 0.47 & 0.46 & 0.40 & 0.42 & 0.26  \\ 
\hline
$   {\P_t\,({\tilde b_1 \to  t\,\tilde \chi^-_1})}$ &   0.92 & $-0.92$ & $-0.56$ & $-0.05$ & 0.99 & $-0.98$ & $-0.94$  \\
$   {\P_t\,({\tilde b_1 \to  t\,\tilde \chi^-_2})}$ & --  & -- &  -- & -- & $-0.29$ & -- & -- \\
$   {\P_t\,({\tilde b_2 \to  t\,\tilde \chi^-_1})}$ &$-0.99$ & 0.99  & $0.95$ &$0.94$& $-0.99$ & 0.99 & $-0.97$  \\
$   {\P_t\,({\tilde b_2 \to  t\,\tilde \chi^-_2})}$ & $-0.94$ & $-0.67$ & -- & -- & $-0.99$ & -- & --  \\
$   {\P_t\,({\tilde t_1 \to  t\,\tilde \chi^0_1})}$ &0.82  & 0.40 & 0.88 &0.86& 0.99 & 0.94 & $-0.99$  \\
$   {\P_t\,({\tilde t_1 \to  t\,\tilde \chi^0_2})}$ &0.92 & 0.50& 0.97 &$0.96$& 0.99 & 0.99 & $-0.95$  \\ 
\hline
$   {\widehat\P_t\,({\rm total})}$ & 0.75 & $-0.73$ &0.46 & 0.64 &0.92 & 0.07 & $-0.83$  \\
$   {A_{\theta_l}}$ & 0.59 & 0.14 & 0.61 & 0.60 & 0.80  & 0.47 & 0.12 \\
$   {A_{\phi_l}}$     & 0.84 & 0.57 & 0.83 & 0.84 & 0.92 & 0.76 & 0.55  \\
\hline
\end{tabular}
\caption{Production cross sections at $\sqrt s=14$~TeV, decay branching ratios, top polarizations originating from different decay processes 
in the respective squark rest frame, total polarization in the laboratory frame, as well as polar and azimuthal angle asymmetries for the benchmark points defined in Table~\ref{tab:bm}. 
\label{tab:net_pol}}
\end{table}

Since the dependence of the top polarization on the sbottom and stop masses and mixings is most 
interesting for the higgsino scenario, most of our examples focus on this case. Concretely, points 
{\tt bm-1} to {\tt bm-4} and {\tt bm-6} have $\mu=350$~GeV and $M_2=2M_1=1000$~GeV, 
as used earlier in this paper, leading to a 97\% higgsino LSP and a $\tilde\chi_1^\pm - \tilde\chi_1^0$ 
mass difference of about 9 GeV;  {\tt bm-5} has a similar electroweak-ino pattern but for a lighter mass scale. 
The case of $\tilde\chi_1^\pm\sim \widetilde W^\pm$ is exemplified in {\tt bm-7}.  
For all points, the parameters in the squark sector are adjusted such that $m_{\sbottom_1}\approx 650$~GeV 
(to avoid kinematic effects on $\P_t$) and $m_h\approx126$~GeV. 
All our benchmark points are for $\tan\beta=10$; we do not present specific points for large 
$\tan\beta$ because they would not add any new features w.r.t.\ points  {\tt bm-1} to {\tt bm-7}.
Note also that we have chosen our benchmark points such that they lie (just) outside the current 
exclusion limits---prospects should be good to test them early in the next phase of LHC running 
at 13--14~TeV. The detailed characteristics of the various benchmark points are as follows:

\begin{itemize}

\item {\tt bm-1} features a higgsino scenario with an almost pure LH $\sbottom_1$, for which 
we expect a top polarization in $\sbottom_1\to t\,\tilde\chi_1^-$ decays ($\rm BR\sim 96\%$) close to $+1$. 
As the light sbottom mass is determined by $M_{\tilde Q_3}=571$~GeV, also the $\stop_1$ is light; in fact 
it is 20~GeV lighter than the $\sbottom_1$ and decays with 96\% BR to $t\,\tilde\chi_{1,2}^0$, thus adding to 
the signal and the net polarization of interest. The $\sbottom_2$ and $\stop_2$ are heavy and play 
no role for our analysis. The tops stemming from $\stop_1$ decays also have a polarization close 
to $+1$, so we expect a large positive net polarization, diluted only by boost effects. Indeed, we find 
$\widehat\P_t=0.75$ for this scenario, together with rather large asymmetries ${A_{\theta_l}}\approx0.6$ 
and ${A_{\phi_l}}\approx0.8$.
 
\item {\tt bm-2} is an example of the higgsino scenario with an almost pure RH $\sbottom_1$, 
leading to a top polarization close to $-1$ in $\sbottom_1\to t\,\tilde\chi_1^-$ decays. 
The $\stop_1$ can be chosen heavy in this case, so that the only relevant cross section is 
$\sbottom_1^{}\sbottom_1^*$ production. Compared to {\tt bm-1}, the total signal is further reduced
by the smaller BR($\sbottom_1\to t\,\tilde\chi_1^-$) of only 34\%. In the lab frame, we have 
$\widehat\P_t=-0.73$ for this scenario, with a small polar angle asymmetry of ${A_{\theta_l}}\approx0.1$
but a still sizable azimuthal angle asymmetry of ${A_{\phi_l}}\approx0.6$.

\item {\tt bm-3} and {\tt bm-4} illustrate the case of intermediate polarization, which is achieved by giving 
the mostly RH $\sbottom_1$ a small  $\sbottom_L$ component. 
Point {\tt bm-3} has $\cos\theta_{\sbottom}=0.1$ and ${\P_t\,({\tilde b_1 \to  t\,\tilde \chi^-_1})}=-0.5$, 
while {\tt bm-4} has $\cos\theta_{\sbottom}=0.15$ and ${\P_t\,({\tilde b_1 \to  t\,\tilde \chi^-_1})}\approx 0$.
In contrast to the two previous points, both $\sbottom_2$ and $\stop_1$ have masses around 700~GeV 
({\it i.e.}\ not far from the $\sbottom_1$) and thus also significantly contribute, in particular because both  
BR($\sbottom_2\to t\,\tilde\chi_1^-$) and ${\rm BR}(\tilde t_1 \to  t\,\tilde \chi^0_{1,2})$ are very large, 
around 96\%.  
The net polarization in the lab frame turns out to be $0.46$ ($0.64$) for {\tt bm-3} ({\tt bm-4}).  
The polar and azimuthal angle asymmetries are sizable, ${A_{\theta_l}}\approx0.6$ and ${A_{\phi_l}}\approx0.8$, 
for both points. 
Note however that with roughly 210--250~fb the total rate ($\sigma\times$BR's) before cuts  
for $t\bar t + E_T^{\rm miss}$ is smaller than for  {\tt bm-1} (roughly 290~fb). 

\item {\tt bm-5} is an example for the situation when both charginos are light enough to allow 
$\tilde b_1 \to  t\,\tilde \chi^-_1$ and $\tilde b_1 \to  t\,\tilde \chi^-_2$ decays. As before, 
the LSP is mostly higgsino and the $\tilde\chi_1^\pm-\tilde\chi_1^0$ mass difference is small, such that the 
$\tilde\chi_1^\pm\to \tilde\chi_1^0\, W^*$ is probably missed. The $\tilde\chi_2^\pm$ on the other hand is wino-like 
and decays to $\tilde\chi_1^0$ plus an on-shell $W$. The top polarization in the sbottom rest frame is $\approx+1$ 
for $\sbottom_1$ decays into $t\,\tilde \chi^-_1$, while it has a small negative value, $\P_t=-0.29$, 
for the decays into  $t\,\tilde \chi^-_2$. 
As for {\tt bm-1}, the $\stop_1$ is also light and decays with $\sim67\%$ to $t\, \tilde\chi_{1,2}^0$. 
The top polarization in these decays, which can likely not be distinguished from $\tilde b_1 \to  t\,\tilde \chi^-_1$, 
is close to $+1$. We end up with $\widehat\P_t=+0.92$ in the lab frame and large 
asymmetries of ${A_{\theta_l}}\approx0.8$ and ${A_{\phi_l}}\approx0.9$.

\item {\tt bm-6} is a variant of {\tt bm-3} with somewhat heavier $\stop_1$ and $\sbottom_2$. 
It is constructed such that the net polarization resulting from the combination of 
${\P_t\,({\tilde b_1 \to  t\,\tilde \chi^-_1})}\approx+1$ and 
${\P_t\,({\tilde t_1 \to  t\,\tilde \chi^0_{1,2}})}\approx-1$ is almost zero. 
Nonetheless the asymmetry parameters are sizable, ${A_{\theta_l}}\approx0.5$ and ${A_{\phi_l}}\approx0.8$. 

\item {\tt bm-7} shows the wino-like $\tilde \chi^\pm_1$ case, which gives $\P_t\approx -1$ in sbottom decays 
whatever is the sbottom mixing angle. Note that the $\stop_1$ decays into $t\,\tilde\chi_1^0$ or 
$t\,\tilde\chi_2^0$ also give $\P_t\approx -1$. However, these decays can in principle be distinguished, 
as  $m_{\tilde\chi_1^0}\simeq250$~GeV and $\tilde\chi_2^0\to Z\tilde\chi_1^0$ and 
$\tilde\chi_1^\pm\to W\tilde\chi_1^0$. To estimate $\widehat\P_t$ and the leptonic asymmetries, 
we sum over $\sbottom_1$ and $\sbottom_2$ production followed by decays into $t\,\tilde \chi^\pm_1$. 
We find $\widehat\P_t=-0.83$, ${A_{\theta_l}}\approx0.1$ and ${A_{\phi_l}}\approx0.5$. 

\end{itemize}

\subsection{\mbf Polarization dependent kinematic distributions} \label{sect:pol_observables}

In this section we present results for the polarization-dependent kinematic distributions of the top decay products 
introduced in Section~\ref{sect:observables}. 
As mentioned, these distributions depend on the top polarization in the sfermion rest frame, the top boost in the laboratory frame and the transverse momentum of the top quark. As a result they are influenced by the mass difference between the decaying sfermion and the daughter electroweak-ino as discussed in ref.~\cite{Belanger:2012tm}. Here, we concentrate on the case of 
large mass difference. More concretely, we analyze the lepton energy distributions as well as the energy ratios  $z$ and $u$ 
of eq.~(\ref{zudef}) for  {\tt bm-2}, {\tt bm-5} and {\tt bm-6}, corresponding to close to $\pm1$ and zero net polarization, respectively.

\begin{figure}[t]\centering
    \hspace*{-3mm}\includegraphics[width=0.54\textwidth]{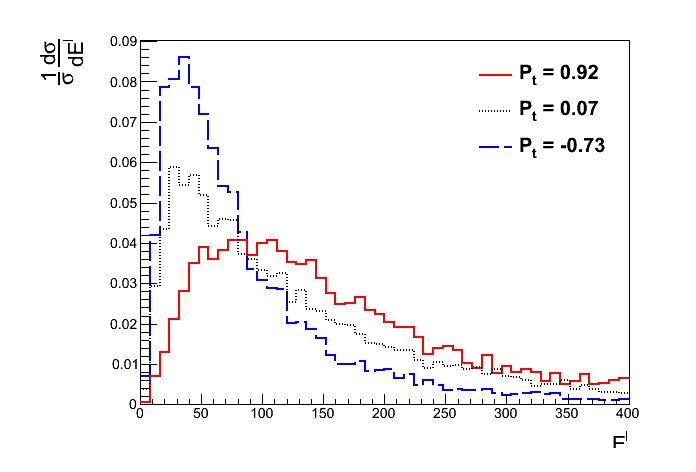}\hspace*{-6mm}\includegraphics[width=0.54\textwidth]{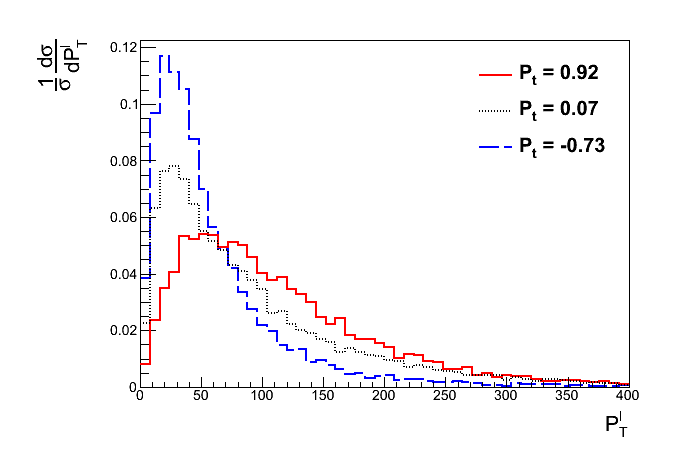}
   \vspace*{-6mm}   
\caption{Lepton energy $E_l$ (left) and transverse momentum $p_T^{\,l}$ (right) distributions for three 
different polarizations. 
The dashed blue lines are for {\tt bm-2} with $\widehat \P_t=-0.73$, the full red lines are for {\tt bm-5} with 
$\widehat \P_t=0.92$, and the dotted black lines are for {\tt bm-6} with $\widehat \P_t=0.07$. 
\label{fig:obs_lep_dist}}
\end{figure}

Figure~\ref{fig:obs_lep_dist} shows the lepton energy $E_l$ (left) and transverse momentum $p_T^{\,l}$ (right) 
distributions for {\tt bm-2}, {\tt bm-5} and {\tt bm-6}.  
As can be seen from eq.~(\ref{eq:topdecay})  for a positively polarized top, the decay leptons go preferentially 
in the forward direction and hence, including the boost, the energy and transverse momentum of the leptons 
are larger than in the unpolarized or negatively polarized case. 
For the negatively polarized top, the situation is opposite, the leptons gain less energy and the distributions 
peak at smaller values. 
This effect is clearly seen in both the $E_l$ and $p_T^{\,l}$ distributions in Fig.~\ref{fig:obs_lep_dist} when 
comparing the lines  for $\P_t = +0.92$, $0.07$ and $-0.73$. 

We conclude that for $\sbottom_1\sim\sbottom_L$ decaying into a higgsino-like chargino 
(and/or $\stop_1\sim\stop_L$ decaying into top+neutralino~\cite{Belanger:2012tm}), the harder 
$E_l$ and $p_T^{\,l}$ distributions can be used to extend 
the reach of searches for 3rd generation squarks. On the other hand, for scenarios which lead to small or negative 
top polarization, the opposite conclusion holds.  Note, however, that the polarization-dependence of  the $E_l$ and $p_T^{\,l}$ distributions makes a general interpretation in terms of SMS \cite{Okawa:2011xg,Chatrchyan:2013sza} 
difficult  if information on the polarization is used in the analysis.

\begin{figure}[t!]\centering
    \hspace*{-3mm}\includegraphics[width=0.54\textwidth]{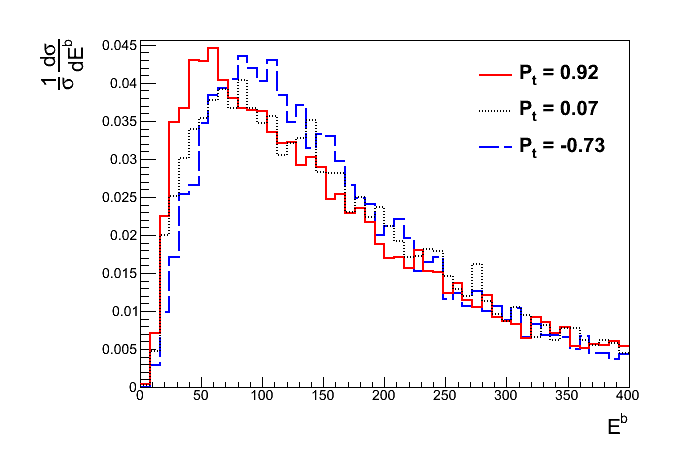}\hspace*{-6mm}\includegraphics[width=0.54\textwidth]{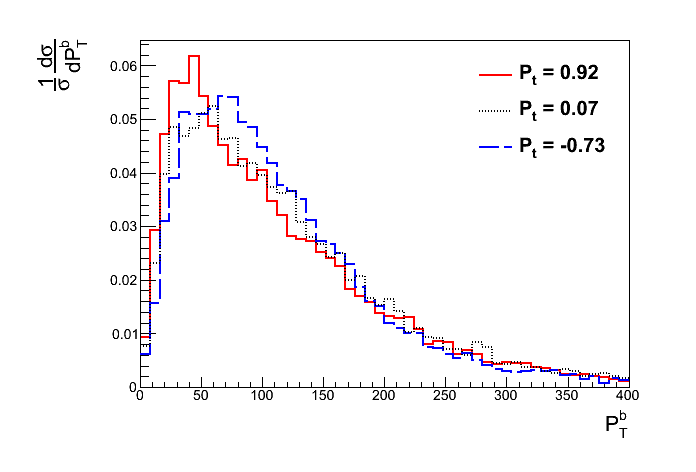}
   \vspace*{-6mm}   
\caption{Same as Fig.~\ref{fig:obs_lep_dist} but for 
$b$--jet energy $E^{b}$ (left) and transverse momentum $p_T^{\,b}$ (right) distributions. 
\label{fig:obs_b_dist}}
\end{figure}

The behavior of the $E_{l}$ and $p_{T}^{\ l}$ distributions
for different polarizations seen in Fig.~\ref{fig:obs_lep_dist}
is a reflection of the angular distribution of the lepton w.r.t.\ the top quark spin direction 
in the rest frame given by eq.~(\ref{eq:topdecay}). Since  $\kappa_{l}$ and $\kappa_{b}$
have opposite signs, the corresponding distributions for $b$-jets will show the opposite behavior. 
Figure~\ref{fig:obs_b_dist} shows the $b$--jet energy $E^{b}$ (left) and transverse momentum $p_T^{\,b}$ (right) 
distributions for {\tt bm-2}, {\tt bm-5} and {\tt bm-6}. 
Although the size of the effect is smaller then for leptons due to the smaller value of $\kappa_{b}$, 
the $b$-jet distributions may still provide interesting complementary information.
In particular, the distributions get harder as the top polarization changes from $+1$ to $-1$. 
Thus the loss of reach due to a softened lepton spectrum in case of a negatively polarized top
might be compensated to some extent by the harder $b$-jet spectrum.  

\begin{figure}[t]\centering
   \hspace*{-3mm}\includegraphics[width=0.54\textwidth]{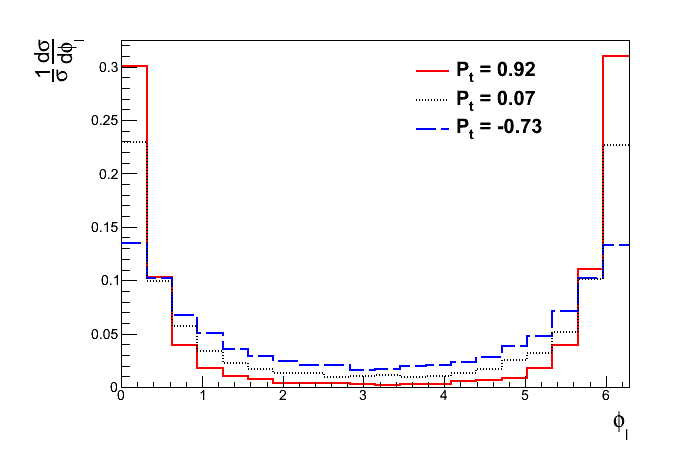}\hspace*{-6mm}\includegraphics[width=0.54\textwidth]{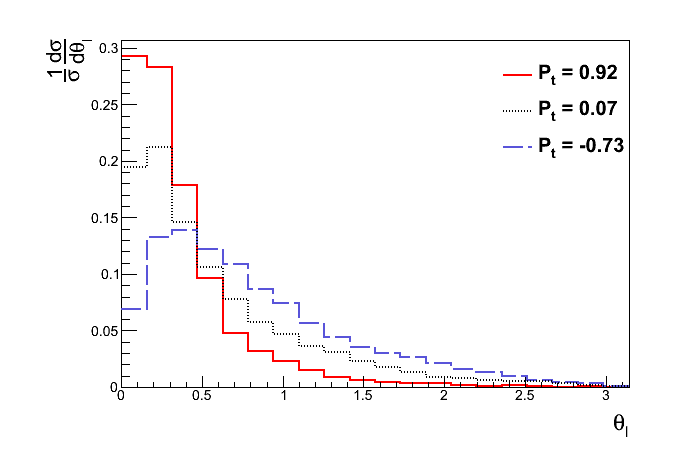}
   \vspace*{-6mm}   
\caption{Distributions of the azimuthal angle  $\phi_l$ (left) and the polar angle 
$\theta_l$ (right) of the decay lepton. The dashed blue lines are for {\tt bm-2} with $\widehat \P_t=-0.73$, 
the full red lines are for {\tt bm-5} with $\widehat \P_t=0.92$, and the dotted black lines are for {\tt bm-6} 
with $\widehat \P_t=0.07$. 
\label{fig:obs_asymmetry}}
\end{figure}

In Section~\ref{sect:observables} we also discussed that the azimuthal angle asymmetry $A_{\phi_l}$ and the 
polar angle asymmetry $A_{\theta_l}$ may also give us a quantitive measure for the top polarization. 
To illustrate this point, we show in Fig.~\ref{fig:obs_asymmetry} the azimuthal angle $\phi_l$ (left) and 
polar angle $\theta_l$ (right) distributions of the decay leptons for {\tt bm-2}, {\tt bm-5} and {\tt bm-6}.
As expected, the distributions for $\phi_l$ peak at $\phi_l=0$ and $\phi_l=2\pi$ for all the three cases, 
but the peaks are higher for a positively polarized tops as compared to unpolarized or the negatively polarized ones. 
A similar situation is seen for the distribution of the polar angle $\theta_l$: the peaking is again in the direction 
of the top boost and increases when going from negative to positive polarization.
The specific values of the asymmetries $A_{\theta_l}$ and $A_{\phi_l}$, defined by eqs.~(\ref{athetadef}) and (\ref{Aldef}), are given in Table~\ref{tab:net_pol}.
The values of both the asymmetries are the lowest for negative polarization and increase as the polarization goes to $1$, thus making them a measure of the polarization.

\begin{figure}[t]\centering
   \includegraphics[width=0.54\textwidth]{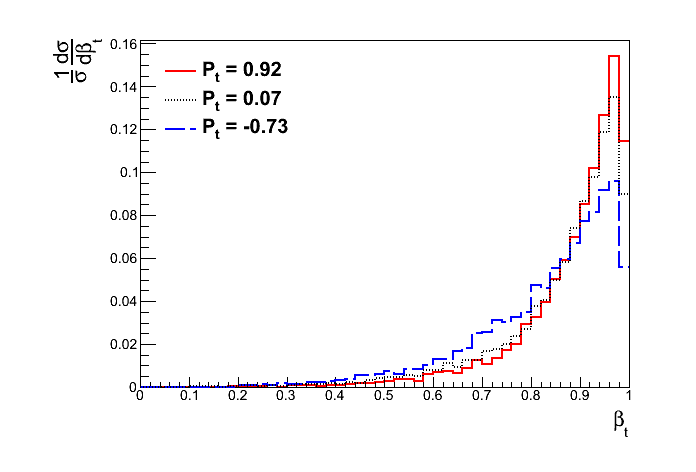}
   \vspace*{-4mm}   
\caption{The top boost distribution at 14~TeV,  for {\tt bm-2} with $\widehat \P_t=-0.73$ (dashed blue line), 
{\tt bm-5} with $\widehat \P_t=0.92$ (full red line), and {\tt bm-6} with $\widehat \P_t=0.07$ (dotted black line). 
\label{fig:obs_topboost}}
\end{figure}

The above observables concerning angular distributions of the leptons are independent of the anomalous $tbW$ coupling. 
At LHC-14, however, the tops will be highly boosted and it may not be easy to use the angular observables. The boost distribution of the tops for the three benchmark points under consideration is shown in Fig.~\ref{fig:obs_topboost}. 
The boost is of course independent of the top polarization---the small differences in the boost distributions arise from the 
different stop and sbottom masses for the three benchmark scenarios. The main point is that the tops are typically 
highly boosted. In such a situation, the energy ratios $u$ and $z$ defined in eq.~(\ref{zudef}) may give very useful information. 
(Recall that these  observables can be however affected by nonzero values of an anomalous $tbW$ coupling.) 
A cut on the top boost $\beta_t$ can enhance the dependence on the polarization.

\begin{figure}[t]\centering
   \includegraphics[width=0.54\textwidth]{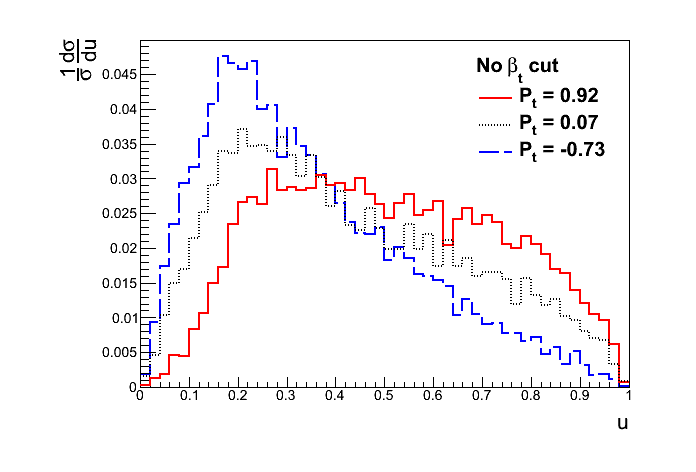}\\
   \hspace*{-4mm}\includegraphics[width=0.54\textwidth]{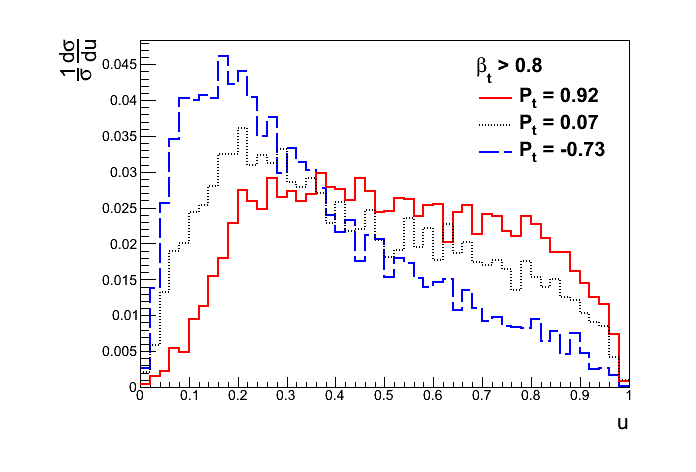}\hspace*{-6mm}\includegraphics[width=0.54\textwidth]{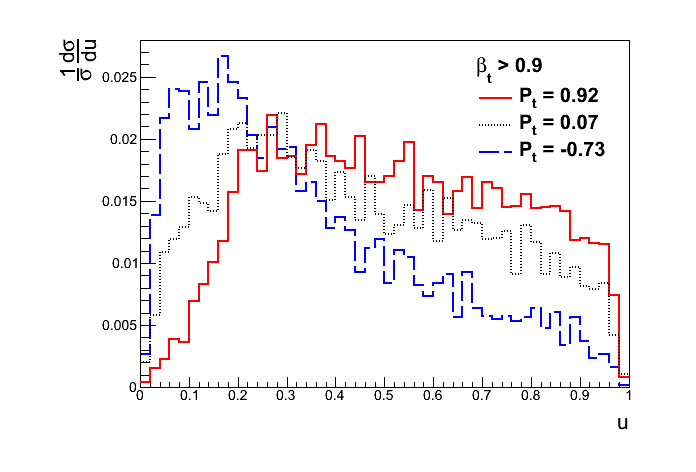}
\vspace*{-6mm}   
\caption{Distributions of the energy ratio $u$, without cut on the boost (top), for $\beta_t > 0.8$ (lower left) 
and for $\beta_t > 0.9$ (lower right).
The dashed blue lines are for {\tt bm-2} with $\widehat \P_t=-0.73$, the full red lines are for {\tt bm-5} with 
$\widehat \P_t=0.92$, and the dotted black lines are for {\tt bm-6} with $\widehat \P_t=0.07$. 
The distributions show a strong dependence on the values of polarization considered. \label{fig:obs_lep_ratio_u}}
\end{figure}

\begin{figure}[t]\centering
   \includegraphics[width=0.54\textwidth]{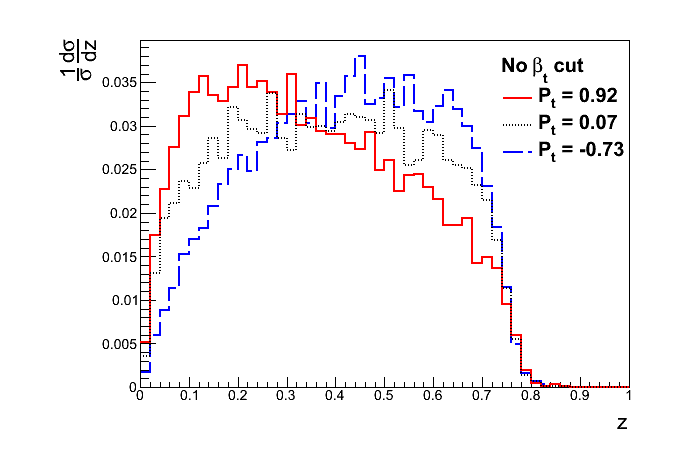}\\
   \hspace*{-4mm}\includegraphics[width=0.54\textwidth]{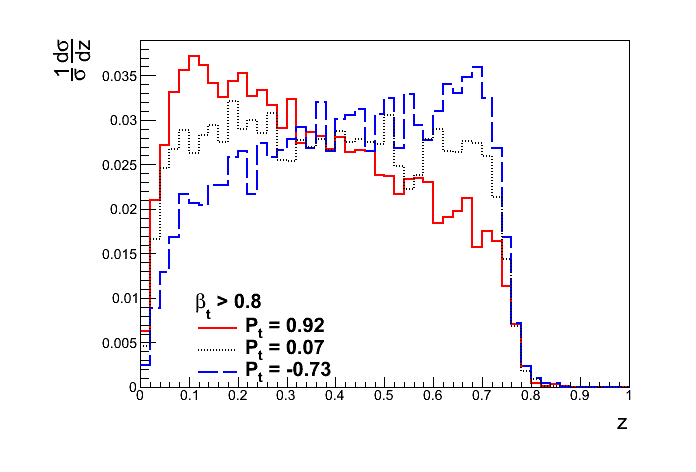}\hspace*{-6mm}\includegraphics[width=0.54\textwidth]{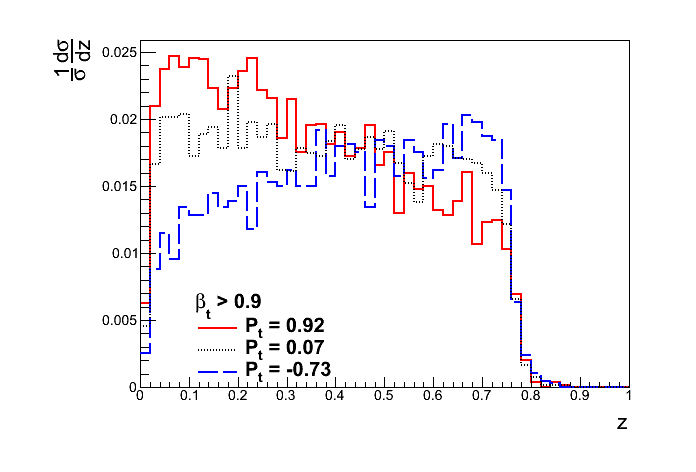}
\vspace*{-6mm}   
\caption{Distributions of the energy ratio $z$, without cut on the boost (top), for $\beta_t > 0.8$ (lower left) 
and for $\beta_t > 0.9$ (lower right). 
The dashed blue lines are for {\tt bm-2} with $\widehat \P_t=-0.73$, the full red lines are for {\tt bm-5} with 
$\widehat \P_t=0.92$, and the dotted black lines are for {\tt bm-} with $\widehat \P_t=0.07$. 
The distributions show a strong dependence on the values of polarization considered.
\label{fig:obs_lep_eratio}}
\end{figure}

Figure~\ref{fig:obs_lep_ratio_u} shows the distributions of the energy ratio $u$ for different cuts on $\beta_t$. 
The distributions are weighted towards smaller values of $u$ for negatively polarized tops, and towards higher values of $u$ for a positive polarization. The cut on the boost enhances the separation of +ve and $-$ve polarization. 
The analogous distributions of the energy ratio $z$ are shown in Fig.~\ref{fig:obs_lep_eratio}. 
As expected from the discussion in Section~\ref{sect:observables}, the behavior of $z$ is opposite to that of $u$, that is positive polarization favours low $z$ values.  
A cut on $\beta_t$ again helps to better differentiate between different values of polarization.  
Clearly, asymmetries similar to $A_{\theta_l}$ and $A_{\phi_l}$ can be constructed for $u$ and $ z$ 
and may serve as an additional measure of the polarization.

\section{Conclusions}

Given the increasingly strong limits on squark and gluino masses from searches at the LHC, together with a  
SM-like Higgs boson with $m_h\approx 125$~GeV, ``natural SUSY'' with 
light 3rd generation squarks, maximal stop mixing, and perhaps light higgsinos emerges 
as the new phenomenological paradigm for SUSY phenomenology to be addressed by the 
next phase of LHC running at 13--14 TeV.  
It is therefore particularly interesting and timely to develop methods to 
a) enhance the discovery potential for stops and sbottoms and
b) determine their properties at the LHC. 

In this paper we discussed the polarization of top quarks stemming from sbottom and stop decays 
as a useful tool to this end.  
In particular we investigated in detail the behavior of the top polarization in sbottom  decays into 
charginos, ${\tilde b}\rightarrow t \tilde\chi^-$. As in the case of stop decays to $t \tilde\chi^0$, 
this polarization may give clues to the nature of both the sbottom and the chargino. 
Concretly, we pointed out that for a mostly wino-like chargino, the top polarization is always $\approx -1$, 
while for a higgsino-like chargino it can take any value between $-1$ and $+1$, depending on the sobttom mixing angle.
Moreover, we discussed that in realistic setups the relation between top polarization and underlying MSSM scenario may not be straightforward because the tops may come from several different channels.  
For example, when the lighter sbottom $\sbottom_1$ has a large LH component, its mass is similar to that of the lighter stop $\stop_1$ (as both are determined by the same soft mass parameter), and therefore both stop 
and sbottom decays may  lead to the same signature, and only the net polarization resulting from all 
different decays will be measurable. 

We illustrated the relation between the top polarizations 
from specific decay channels (in the respective stop or sbottom rest frame) and the net polarization 
in the laboratory by means of seven benchmark points.
Furthermore, we studied how the top polarization affects its  decay kinematics and pointed out the strong correlation between the polarization of the top quark and the kinematical distributions of the various decay products. 
We observed how these  may be used to enhance and evaluate the discovery 
reach, as well as to construct measures of this
polarization using the angular observables of the decay lepton. 

In summary, we have shown that top polarization may provide a useful tool for the searches 
for 3rd generation squarks, in particular in the context  of natural SUSY with light higgsinos, 
a scenario which is very difficult to resolve at the LHC. 

\clearpage

\section*{Acknowledgements}

We thank I.~Niessen and A.~Pukhov for help with technical questions and Dan Tovey for useful discussions. 
This work was partly funded by the French ANR project DMAstroLHC.
GB and SK$\times2$ also  thank the Galileo Galilei Institute for Theoretical Physics (GGI Florence) 
for hospitality and the INFN for partial support. RG wishes to thank the theory group at LPSC (Grenoble) for hospitality when part of this work was done. In addition she would like to acknowledge support from the Department of Science and Technology, 
India  under the J.C. Bose Fellowship  scheme under grant no. SR/S2/JCB-64/2007.

\clearpage

\providecommand{\href}[2]{#2}\begingroup\raggedright\endgroup

\end{document}